\documentclass[aps,preprint]{revtex4}
\usepackage{amsmath}
\usepackage[english]{babel}
\usepackage[latin1]{inputenc}
\usepackage{amssymb}
\usepackage{graphicx}
\usepackage{natbib}
\usepackage{epsfig}
\usepackage{bm}
\usepackage{dcolumn}
\usepackage{color}

\def\tra{{\rm Tr}\,}
\def\trmin{{\rm tr}}
\def\mfa{{\mbox{\tiny MFA}}}

\begin{document}

\vspace*{2cm}

\title{\sc\Large{Properties of magnetized neutral pions at zero and finite temperature
in nonlocal chiral quark models}}

\author{D. G\'omez Dumm$^{a,b}$, M.F. Izzo Villafa\~ne$^{b,c}$, N.N.\ Scoccola$^{b,c}$}

\affiliation{$^{a}$ IFLP, CONICET $-$ Dpto.\ de F\'{\i}sica, Fac.\ de Cs.\ Exactas,
Universidad Nacional de La Plata, C.C. 67, (1900) La Plata, Argentina}
\affiliation{$^{b}$ CONICET, Rivadavia 1917, (1033) Buenos Aires, Argentina}
\affiliation{$^{c}$ Physics Department, Comisi\'{o}n Nacional de Energ\'{\i}a At\'{o}mica, }
\affiliation{Av.\ Libertador 8250, (1429) Buenos Aires, Argentina}

\begin{abstract}
The behavior of $\pi^0$ meson properties in the presence of a uniform
external magnetic field is studied in the context of a nonlocal extension of
the Polyakov-Nambu-Jona-Lasinio model. The analysis includes the $\pi^0$
mass, the effective $\pi^0$-quark coupling and the pion-to-vacuum hadronic
form factors, both at zero and finite temperature. Numerical results are
compared with previous calculations carried out within the local NJL model,
when available. The validity of chiral relations and the features of
deconfinement and chiral symmetry restoration transitions are discussed.
\end{abstract}
%\date{\today}

%\pacs{21.65.Qr, 25.75.Nq, 75.30.Kz, 11.30.Rd}

\maketitle

\renewcommand{\thefootnote}{\arabic{footnote}}
\setcounter{footnote}{0}

\section{Introduction}

The behavior of strongly interacting matter under the influence of intense
magnetic fields has become an issue of increasing interest in the last few
years~\cite{Kharzeev:2012ph,Andersen:2014xxa,Miransky:2015ava}. This is
mostly motivated by the realization that the presence of strong magnetic
fields should be taken into account in the analysis of some relevant
physical systems, e.g.\ in the description of high energy non-central heavy
ion collisions~\cite{HIC}, or the study of magnetars~\cite{duncan}. From the
theoretical point of view, addressing this subject requires to deal with
quantum chromodynamics (QCD) in nonperturbative regimes. Therefore, present
analyses are based either in the predictions of effective models or in the
results obtained through lattice QCD (LQCD) calculations. In this work we
focus on the effect of an intense external magnetic field on various $\pi^0$
meson properties at zero and finite temperature. This issue has been studied
in the last years following various theoretical approaches for low-energy
QCD, such as Nambu-Jona-Lasinio (NJL)-like
models~\cite{Fayazbakhsh:2013cha,Fayazbakhsh:2012vr,Avancini:2015ady,Zhang:2016qrl,Avancini:2016fgq,Mao:2017wmq,Avancini:2018svs},
chiral perturbation theory (ChPT)~\cite{Andersen:2012zc,Agasian:2001ym} and
path integral Hamiltonians
(PIH)~\cite{Orlovsky:2013wjd,Andreichikov:2016ayj}. In addition, results for
the light meson spectrum under background magnetic fields at zero
temperature have been obtained from LQCD
calculations~\cite{Bali:2015vua,Bali:2017ian}.

In Ref.~\cite{GomezDumm:2017jij} we have studied the behavior of the $\pi^0$
meson mass and one of its axial decay form factors in the presence of a
uniform static magnetic field at zero temperature, within a relativistic
chiral quark model in which quarks interact through a nonlocal four-fermion
coupling~\cite{Rip97}. This so-called ``nonlocal NJL (nlNJL) model'' can be
viewed as a sort of extension of the NJL model that intends to provide a
more realistic effective approach to QCD. Actually, nonlocality arises
naturally in the context of successful descriptions of low-energy quark
dynamics~\cite{Schafer:1996wv,RW94}, and it has been
shown~\cite{Noguera:2008} that nonlocal models can lead to a momentum
dependence in quark propagators that is consistent with LQCD results.
Moreover, in this framework it is possible to obtain an adequate description
of the properties of light mesons in the absence of an external
electromagnetic field at both zero and finite
temperature~\cite{Noguera:2008,Bowler:1994ir,Schmidt:1994di,Golli:1998rf,
General:2000zx,Scarpettini:2003fj,GomezDumm:2006vz,
Contrera:2007wu,Hell:2008cc,Contrera:2009hk,Dumm:2010hh,Carlomagno:2013ona}. Interestingly,
as shown in Refs.~\cite{Pagura:2016pwr,GomezDumm:2017iex}, nlNJL models
naturally allow to reproduce the so-called inverse magnetic catalysis (IMC)
effect, previously observed from LQCD results. According to these
calculations, the chiral restoration critical temperature turns out to be a
decreasing function of the magnetic field $B$. In fact, the observation of
IMC in LQCD calculations~\cite{Bali:2011qj,Bali:2012zg} represents a
challenge from the point of view of theoretical models, since most naive
effective approaches to low energy QCD (NJL model, ChPT, MIT bag model,
quark-meson models) predict that the chiral transition temperature should
grow when the magnetic field is
increased~\cite{Andersen:2014xxa,Kharzeev:2012ph,Miransky:2015ava}. In
addition, nlNJL models lead to a $B$ dependence of the $\pi^0$ mass that is
found to be in good agreement with LQCD results~\cite{GomezDumm:2017jij}.

The aim of the present article is to extend the work in
Ref.~\cite{GomezDumm:2017jij}, considering some additional properties of the
magnetized $\pi^0$ mesons. As shown in Ref.~\cite{Coppola:2018ygv}, in the
presence of a constant magnetic field $\vec B$ the pion-to-vacuum vector and
axial vector amplitudes can be in general parametrized in terms of three
``decay'' form factors. One of them, $f_{\pi^0}^{(A||)}$, corresponds to the
pion decay constant usually denoted by $f_\pi$. The behavior of this form
factor under the magnetic field has already been analyzed in
Ref.~\cite{GomezDumm:2017jij}, together with those of the masses $m_{\pi^0}$
and $m_\sigma$, and the quark-meson coupling $g_{\pi^0qq}$. The other two
decay form factors are a second axial decay constant,
$f_{\pi^0}^{(A\perp)}$, associated to momentum components that are
perpendicular to the magnetic field, and a vector decay constant
$f_{\pi^0}^{(V)}$. As shown in Ref.~\cite{Fayazbakhsh:2013cha}, another
relevant feature induced by the presence of the external magnetic field is
the fact that the $\pi^0$ dispersion relation turns out to be anisotropic,
implying that the movement along the direction perpendicular to the magnetic
field is characterized by a diffraction index $u_{\pi^0}$ which is in
general different from one. In this way, to complement the analysis carried
out in Ref.~\cite{GomezDumm:2017jij}, in this work we study the magnetic
field dependence of $f_{\pi^0}^{(A\perp)}$, $f_{\pi^0}^{(V)}$ and
$u_{\pi^0}$. In addition, we extend the analysis to a system at finite
temperature $T$, considering the thermal behavior of these quantities and
also of the masses $m_{\pi^0}$ and $m_\sigma$, the coupling $g_{\pi^0qq}$
and the decay constant $f_{\pi^0}^{(A||)}$, which have been studied in
Ref.~\cite{GomezDumm:2017jij} only for $T=0$.

This article is organized as follows. In Sec.~II we show how to obtain the
analytical equations required to determine the relevant $\pi^0$ properties
at zero temperature in the presence of the magnetic field. Our calculations
are based on the formalism developed in
Refs.~\cite{Pagura:2016pwr,GomezDumm:2017iex,GomezDumm:2017jij}, which make
use of Ritus eigenfunctions~\cite{Ritus:1978cj}. In Sec.~III we show how to
extend the analysis in Sec.~II to a system at finite temperature, taking
also into account the coupling of fermions to a background color field (the
so-called ``Polyakov loop nlNJL model''). In Sect.~IV we quote and discuss
our numerical results, while in Sec.~V we present our conclusions. Finally,
in Appendices A and B we outline the derivation of some of the expressions
quoted in the main text.

\section{Theoretical formalism}

The Euclidean action for the nonlocal NJL-like two-flavor quark model we are
considering reads
  \begin{equation}
      S_E = \int d^4 x \ \left\{ \bar \psi (x) \left(- i \rlap/\partial +
  m_c \right) \psi (x) \ - \ \frac{G}{2} j_a(x) j_a(x) \right\} \ .
  \label{action}
  \end{equation}
Here $m_c$ is the current quark mass, equal for $u$ and $d$ quarks, while the
currents $j_a(x)$ are given by
  \begin{eqnarray}
    j_a (x) &=& \int d^4 z \  {\cal G}(z) \
    \bar \psi(x+\frac{z}{2}) \ \Gamma_a \ \psi(x-\frac{z}{2}) \ ,
  \label{cuOGE}
  \end{eqnarray}
where $\Gamma_{a}=(\leavevmode\hbox{\small1\kern-3.8pt\normalsize1},i\gamma
_{5}\vec{\tau})$. The function ${\cal G}(z)$ is a nonlocal form factor that
characterizes the effective interaction. The action can be ``gauged'' to
incorporate couplings to electromagnetic, vector and axial vector gauge
fields $\mathcal{A}_{\mu}$, $W_\mu^{V,a}(x)$ and $W_\mu^{A,a}(x)$,
respectively. This is done by replacing
\begin{equation}
    \partial_{\mu}\ \rightarrow\ D_\mu\equiv\partial_{\mu}-i\,\hat Q \mathcal{A}_{\mu}(x)\
            - \ \dfrac{i}{2}\,\Gamma^C\, \tau^a\, W_\mu^{C,a}(x) \ ,
\label{covdev}
\end{equation}
where $\hat Q=\mbox{diag}(q_u,q_d)$, with $q_u=2e/3$, $q_d = -e/3$, $C=V,A$,
$a=1,2,3$, $\Gamma^V=1$ and $\Gamma^A=\gamma_5$. For this nonlocal model,
gauge symmetry also requires the
replacements~\cite{GomezDumm:2006vz,Noguera:2008,Dumm:2010hh}
\begin{equation}
    \psi(x-z/2) \rightarrow \mathcal{W}\left(  x,x-z/2\right)  \, \psi(x-z/2)\
    ,\quad
    \psi(x+z/2)^\dagger \rightarrow \psi(x+z/2)^\dagger\,\mathcal{W}\left(  x+z/2,x\right) \
    ,
\label{transport}
\end{equation}
with
\begin{equation}
   \mathcal{W}(x,y) = \exp\left[ -i \left( \hat Q \int_x^y d\ell_\mu A_\mu(\ell) +
      \frac{\tau^a}{2} \Gamma^C \int_x^y d\ell_\mu \ W^{C,a}_\mu(\ell) \right)
      \right]\ ,
\label{intpath}%
\end{equation}
where $\ell$ runs over an arbitrary path connecting $x$ with $y$. As it is
usually done, we take it to be a straight line path.

As stated, we assume the presence of an external uniform magnetic field
$\vec B$. Therefore, using the Landau gauge, and choosing the $x_3$ axis in
the direction of $\vec B$, we take $\mathcal{A}_{\mu}$ to be a static field
given by $\mathcal{A}_{\mu}(x) = B x_1\delta_{\mu 2}$.

Since we are interested in studying light meson properties, we carry out a
bosonization of the fermionic theory, introducing scalar and pseudoscalar
fields $\sigma(x)$ and $\vec{\pi}(x)$ and integrating out the fermion
fields. The bosonized action can be written
as~\cite{GomezDumm:2017jij,Noguera:2008,Dumm:2010hh}
\begin{equation}
    S_{\mathrm{bos}} \ = \ -\log\det\mathcal{D} \ + \ \frac{1}{2G}
      \int d^{4}x
      \Big[\sigma(x)\,\sigma(x)+ \vec{\pi}(x)\cdot\vec{\pi}(x)\Big]\ ,
\label{sbos}
\end{equation}
where
\begin{eqnarray}
\mathcal{D}
\left( x,x' \right) &  = & \delta^{(4)}(x-x')\,\big(-i\,\rlap/\!D + m_{c} \big)\,
+ \nonumber \\
& & \mathcal{G}(x-x') \, \gamma_{0} \, {\cal W}(x,\bar x)\,
\gamma_{0} \big[\sigma(\bar x) + i\,\gamma_5\,\vec{\tau}\cdot\vec{\pi}(\bar x) \big]
\, {\cal W}(\bar x,x') \ ,
\label{dxx}
\end{eqnarray}
with $\bar x = (x+x')/2$. We expand now the meson fields around their mean
field values. Since the external magnetic field is uniform, one can assume
that the field $\sigma(x)$ has a nontrivial translational invariant mean
field value $\bar{\sigma}$, while the vacuum expectation values of
pseudoscalar fields are zero. We separate the mean field piece of the first
term of the action in Eq.~(\ref{sbos}), writing
\begin{eqnarray}
    -\log\det\mathcal{D} &=&  - \tra\log\mathcal{D}_0 \, - \,
    \tra\log(1 + \mathcal{D}_0^{-1}\,\delta\mathcal{D})
    \ ,
\label{expansion}
\end{eqnarray}
where the traces run over color, flavor, Dirac and coordinate spaces. The
form of the mean field operator $\mathcal{D}_0$ in the presence of the
external magnetic field has been studied in detail in previous works, see
e.g.~Ref.~\cite{GomezDumm:2017iex}. It can be written as
\begin{equation}
\mathcal{D}_0 \ = \ {\rm diag}\big(\mathcal{D}_u^\mfa(x,x')\, ,\,
      \mathcal{D}_d^\mfa(x,x')\big)\ ,
\end{equation}
where
  \begin{equation}
    \mathcal{D}_f^\mfa(x,x') \ = \ \delta^{(4)}(x-x') \left( - i \rlap/\partial
      - q_f \, B \, x_1 \, \gamma_2 + m_c \right)  + \, \bar\sigma \,
      \mathcal{G}(x-x') \, \exp\left[i\Phi_f(x,x')\right]\ .
  \end{equation}
Here $\Phi_f(x,x')= q_f B \, (x_2 - x_2')\, (x_1 +x_1')/2\,$ is the
so-called Schwinger phase, and a direct product to an identity matrix in
color space is understood. The mean field quark propagators $S_f^\mfa(x,x')
= \big[\mathcal{D}_f^\mfa(x,x')\big]^{-1}$ can be obtained following the
Ritus eigenfunction method~\cite{Ritus:1978cj}. As shown in
Ref.~\cite{GomezDumm:2017iex} (see also the analysis carried out within the
Schwinger-Dyson formalism in Refs.~\cite{Watson:2013ghq,Mueller:2014tea}),
it is possible to write the propagators in terms of the Schwinger phase and a
translational invariant function, namely
  \begin{equation}
    S_f^\mfa(x,x') \ = \ \exp\!\big[i\Phi_f(x,x')\big]\,\int \frac{d^4p}{(2\pi)^4}\
      e^{i\, p\cdot (x-x')}\, \tilde S_f(p_\perp,p_\parallel)\ ,
  \end{equation}
where $p_\perp = (p_1,p_2)$ and $p_\parallel = (p_3,p_4)$. The expression of
$\tilde S_f(p_\perp,p_\parallel)$ in the nlNJL model under consideration is
found to be~\cite{GomezDumm:2017iex}
  \begin{eqnarray}
    \tilde S_f(p_\perp,p_\parallel) &=& 2\, \exp(-p_\perp^2/|q_f B|)
      \sum_{k=0}^\infty \sum_{\lambda=\pm} \Big[(-1)^{k_\lambda}
      \big(\hat A^{\lambda,f}_{k,p_\parallel} - \hat B^{\lambda,f}_{k,p_\parallel}
      \, p_\parallel\cdot\gamma_\parallel\big) L_{k_\lambda}(2p_\perp^2/|q_f B|)
      + \nonumber\\
                    & & 2\, (-1)^k \big(\hat C^{\lambda,f}_{k,p_\parallel}
      - \hat D^{\lambda,f}_{k,p_\parallel}\, p_\parallel\cdot\gamma_\parallel\big)
      \, p_\perp\cdot\gamma_\perp
      \, L^1_{k-1}(2p_\perp^2/|q_f B|)\Big]\,\Delta^\lambda\ ,
  \label{sfp}
  \end{eqnarray}
where the following definitions have been used. The perpendicular and
parallel gamma matrices are collected in vectors $\gamma_\perp =
(\gamma_1,\gamma_2)$ and $\gamma_\parallel = (\gamma_3,\gamma_4)$, while the
matrices $\Delta^\lambda$ are defined as $\Delta^+ = {\rm diag}(1,0,1,0)$
and $\Delta^- = {\rm diag}(0,1,0,1)$. The integers $k_\lambda$ are given by
$k_\pm = k - 1/2 \pm s_f/2$, where $s_f = {\rm sign} (q_f B)$. The functions
$\hat X^{\pm,f}_{k,p_\parallel}$, with $X = A, B, C, D$, are defined as
  \begin{eqnarray}
    \hat A^{\pm,f}_{k,p_\parallel} &=&
      M^{\mp,f}_{k,p_\parallel}\, \hat C^{\pm,f}_{k,p_\parallel} + p_\parallel^2
      \, \hat D^{\pm,f}_{k,p_\parallel}\ ,
      \label{aa} \\
    \hat B^{\pm,f}_{k,p_\parallel} &=& \hat C^{\pm,f}_{k,p_\parallel}
      - M^{\mp,f}_{k,p_\parallel}\, \hat D^{\pm,f}_{k,p_\parallel}\ \ ,
      \label{bb} \\
    \hat C^{\pm,f}_{k,p_\parallel} &=& \frac{2 k |q_f B| + p_\parallel^2 +
      M^{-,f}_{k,p_\parallel} M^{+,f}_{k,p_\parallel}}{\Delta^f_{k,p_\parallel}}
      \ \ ,
      \label{cc} \\
    \hat D^{\pm,f}_{k,p_\parallel} &=& \frac{M^{\pm,f}_{k,p_\parallel}
      - M^{\mp,f}_{k,p_\parallel}}{\Delta^f_{k,p_\parallel}} \ \ ,
      \label{dd}
  \end{eqnarray}
where
  \begin{equation}
    \Delta^f_{k,p_\parallel} = \left( 2 k |q_f B| + p_\parallel^2 +
      M^{+,f}_{k,p_\parallel}\, M^{-,f}_{k,p_\parallel} \right)^2\! +\, p_\parallel^2
      \left( M^{+,f}_{k,p_\parallel} - M^{-,f}_{k,p_\parallel} \right)^2\ .
  \label{delta}
  \end{equation}
The functions $M^{\lambda,f}_{k,p_\parallel}$ play the role of effective
(momentum-dependent) dynamical quark masses in the presence of the magnetic
field. They are given by
  \begin{equation}
      M^{\lambda,f}_{k,p_\parallel} \ = \
      \frac{4\pi}{|q_fB|}\,(-1)^{k_\lambda}
      \int \frac{d^2p_\perp}{(2\pi)^2}\
      M(p_\perp^2 + p_\parallel^2) \,\exp(-p_\perp^2/|q_fB|) \,
      L_{k_\lambda}(2p_\perp^2/|q_fB|)\ ,
  \label{mpk}
  \end{equation}
where
  \begin{equation}
    M(p^2) \ = \ m_c + \bar\sigma\,g(p^2)\ ,
  \end{equation}
$g(p^2)$ being the Fourier transform of the nonlocal form factor ${\cal
G}(x)$. In Eqs.~(\ref{sfp}) and (\ref{mpk}), $L_k(x)$ and $L_k^1(x)$ stand
for generalized Laguerre polynomials, with the convention $L_{-1}(x) =
L_{-1}^1(x) = 0$. The relation in Eq.~(\ref{mpk}) can be understood as a
Laguerre-Fourier transform of the function $M(p^2)$. It is also convenient
to introduce the Laguerre-Fourier transform of the form factor $g(p^2)$,
\begin{equation}
    g^{\lambda,f}_{k,p_\parallel} \ = \
    \frac{4\pi}{|q_fB|}\,(-1)^{k_\lambda}
    \int \frac{d^2p_\perp}{(2\pi)^2}\
    g(p_\perp^2 + p_\parallel^2) \,\exp(-p_\perp^2/|q_fB|) \,
    L_{k_\lambda}(2p_\perp^2/|q_fB|)\ ,
\label{gpk}
\end{equation}
thus one has
  \begin{equation}
    M^{\lambda,f}_{k,p_\parallel} \ = \ \big[1-\delta_{(k_\lambda+1)\,0}\big]
      m_c\, + \,\bar \sigma \, g^{\lambda,f}_{k,p_\parallel}\ .
  \label{mmain}
  \end{equation}
The transform in Eq.~(\ref{gpk}) can be inverted to get
\begin{equation}
    g(p_\perp^2 + p_\parallel^2) \ = \ 2\, e^{-p_\perp^2/|q_fB|}\;
    \sum_{k=0}^\infty\; (-1)^{k_\lambda}\,
    g^{\lambda,f}_{k,p_\parallel}\,L_{k_\lambda}(2p_\perp^2/|q_fB|)\ .
\label{inversa}
\end{equation}

To study the mass and decay form factors of the neutral pion, we expand the
operator $\delta\mathcal{D}(x,x')$ in powers of the meson fluctuations and
the external vector and axial vector fields, keeping up to linear terms in
$\delta \pi_3$, $W_\mu^{V,3}$ and $W_\mu^{A,3}$. We obtain
\begin{equation}
\delta\mathcal{D}(x,x') \ = \  \delta\mathcal{D}_\pi \left( x,x' \right) +
\delta\mathcal{D}_W^{(a)} \left( x,x' \right) + \delta\mathcal{D}_W^{(b)} \left( x,x'
\right) + \delta\mathcal{D}_{W,\pi} \left( x,x'\right)\ , \label{exp}
\end{equation}
where
\begin{eqnarray}
\delta\mathcal{D}_{\pi} \left( x,x' \right) &=& i \gamma_5 \ \tau^0 \ \exp[ i \Phi(x,x')] g(x-x') \ \delta\pi_3(\bar x)\ , \\
\delta\mathcal{D}_W^{(a)} \left( x,x' \right) &=& - \delta^{(4)}(x-x') \, \frac{\tau^3}{2} \sum_{C=V,A}
\bar\Gamma^C \,\gamma_\mu W_\mu^{C,3}( \bar x) \ , \\
\delta\mathcal{D}_W^{(b)} \left( x,x' \right) &=& i \sigma  \frac{\tau^3}{2}\,\exp[i \Phi(x,x')]\,
g(x-x')\sum_{C=V,A} \bar \Gamma^C \left[ U^{C,3}(x,\bar x) - U^{C,3}(\bar    x,x')\right]\ , \\
\delta\mathcal{D}_{W,\pi} \left( x,x' \right) &=& - \frac{1}{2}\exp[ i \Phi(x,x')]
\, g(x-x')\,\times \nonumber \\
& & \sum_{C=V,A} \gamma_5 \, \Gamma^C \left[ U^{C,3}(x,\bar x) - U^{C,3}(\bar x,x')\right] \, \delta\pi_3(\bar
x)\ .
\end{eqnarray}
Here we have used the definitions $\bar x = (x+x')/2$, $\bar \Gamma^C =
\gamma_0\Gamma^C\gamma_0$ and
\begin{equation}
    U^{C,3}(x,y) \ = \ \int_x^y d\ell_\mu \, W^{C,3}_\mu(\ell)\ .
\end{equation}

Given a definite model parametrization, the value of $\bar\sigma$ can be
found by minimization of the effective action at the mean field level. The corresponding
``gap equation'' reads~\cite{GomezDumm:2017iex,Pagura:2016pwr}
\begin{equation}
\frac{\bar \sigma}{G} \ = \ \dfrac{N_C}{\pi}\sum_{f=u,d}
                B_f\sum_{k=0}^\infty \int_{q_\parallel}
                \sum_{\lambda = \pm} g_{k,q_\parallel}^{\lambda,f}
                \,\hat A^{\lambda,f}_{k,p_\parallel}\ .
\label{gapeq}
\end{equation}

\subsection{Pion field redefinition and quark-meson coupling constants}

The calculation of the $\pi^0$ mass in this model has been previously
carried out in Ref.~\cite{GomezDumm:2017jij}. As shown in that paper, the
piece of the bosonized action that is quadratic in the neutral pion fields
can be written as
\begin{eqnarray}
    S_{\rm bos}\big|_{(\delta\pi_3)^2} &=&
    \frac{1}{2}\,\tra(\mathcal{D}_0^{-1}\,\delta\mathcal{D}_\pi)^2\Big|_{(\delta\pi_3)^2}
    + \frac{1}{2G}\int_{t_\perp t_\parallel}\ \delta\pi_3(t)\,\delta\pi_3(-t)
    \nonumber \\
          &=& \frac{1}{2}\int_{t_\perp t_\parallel} \left[F(t_\perp^2,t_\parallel^2) + \frac{1}{G}\right]
    \delta\pi_3(t)\,\delta\pi_3(-t)\ ,
\label{quadact}
\end{eqnarray}
where for integration in two-component momentum spaces we use the notation
\begin{equation}
    \int_{p\, q\, r \dots} \ \equiv \ \int \frac{d^2p}{(2\pi)^2}\; \frac{d^2q}{(2\pi)^2}\; \frac{d^2r}{(2\pi)^2}\ \dots
\end{equation}
Choosing the frame in which the $\pi^0$ meson is at rest, its mass can be
obtained as the solution of the equation
\begin{equation}
    \frac{1}{G} \ + \ F(0,-m_{\pi^0}^2) \ = \ 0\ .
\label{pimass}
\end{equation}

To normalize the pion field we can expand the action in Eq.~(\ref{quadact})
around the pion pole ($t_\perp=0$, $t_\parallel^2=-m_{\pi^0}^2$) up to first
order in momentum squared. We define
\begin{eqnarray}
    Z_{\parallel}^{-1}  & = & \dfrac{dF(t_\perp^2,t_\parallel^2)}{dt_\parallel^2}
      \bigg\rvert_{t_\perp^2 = 0,\,t_\parallel^2=-m_{\pi^0}^2} \ %\equiv \ g_{\pi^0qq}^{-2} \
      , \nonumber \\
    Z_{\perp}^{-1} & = & \dfrac{dF(t_\perp^2,t_\parallel^2)}{dt_\perp^2}
          \bigg\rvert_{t_\perp^2 = 0,\,t_\parallel^2=-m_{\pi^0}^2} \ , \label{defzperp}
\end{eqnarray}
and renormalize the pion field according to $\pi_3(q)= g_{\pi^0 q q} \,
\tilde{\pi}_3(q)$, where $g_{\pi^0 q q}=Z_{\parallel}^{1/2}$ is the
meson-quark effective coupling constant. Thus, one has
\begin{equation}
    S^{\,\mbox{\tiny quad}}_{\pi^0} \ = \ \dfrac{1}{2} \int_{q_\perp q_\parallel}
        \delta\tilde{\pi}_3(-q) \, \Big( u_{\pi^0}^2\, q_\perp^2 +
        q_\parallel^2+m_{\pi^0}^2 \Big)\, \delta\tilde{\pi}_3(q) \ ,
\label{actionquadpi0p_2}
\end{equation}
where
\begin{equation}
u_{\pi^0}^2 \ = \ \dfrac{Z_{\parallel}}{Z_{\perp}}\ .
\label{u_pi}
\end{equation}

{}From the above expressions of the quark propagators and
$\delta\mathcal{D}_\pi$, after some straightforward calculation we find
\begin{eqnarray}
    F(t_\perp^2,t_\parallel^2) &=& - 16\,\pi^2\,N_C\, \sum_{f=u,d}\,\frac{1}{(q_f B)^2}
                    \int_{q_\perp\, p_\perp\, p'_\perp\, q_\parallel} g(q_\perp^2 +
                    q_\parallel^2)\, g[(p'_\perp + p_\perp - q_\perp)^2\! + q_\parallel^2]\,\times
    \nonumber \\
                   & & \exp[i2\phi(q_\perp,p_\perp,p'_\perp,t_\perp)/(q_f B)] \ {\rm tr}_D \Big[
                    \tilde S_f(p_\perp,q_\parallel^+)\,\gamma_5\,
                    \tilde S_f(p'_\perp,q_\parallel^-)\,\gamma_5\Big]\ ,
\label{f0k}
\end{eqnarray}
where the trace is taken over Dirac space. We have defined $q_\parallel^\pm
= q_\parallel \pm t_\parallel/2$, while the function $\phi$ in the
exponential is given by
\begin{eqnarray}
    \phi(q_\perp,p_\perp,p'_\perp,t_\perp) \ &=& \ p_2\,p'_1 + q_1\,(p'_2 - p_2)
    - p_1\,p'_2 - q_2\,(p'_1 - p_1) \nonumber \\
    && +\, t_2 (q_1-(p_1+p'_1)/2) - t_1 (q_2-(p_2+p'_2)/2) \ . % \nonumber \\
%    &=& \left( q_\perp - (p'_\perp + p_\perp)/2\right) \times \left( t_\perp + p'_\perp - p_\perp\right)
\label{phi}
\end{eqnarray}
As stated in Ref.~\cite{GomezDumm:2017jij}, the trace in Eq.~(\ref{f0k}) is
given by
\begin{eqnarray}
\trmin_D \Big[ \tilde S_f(p_\perp,q_\parallel^+)\,\gamma_5\,
           \tilde S_f(p'_\perp,q_\parallel^-)\,\gamma_5\Big] &=&
      8\,e^{-(p_\perp^2+{p'_\perp}^{\!\! 2})/B_f}\sum_{k,k'=0}^\infty
      (-1)^{k+k'}\,\times \nonumber \\
            & & \hspace{-1.4cm} \bigg[\sum_{\lambda = \pm}
      F_{kk',q_\parallel^+q_\parallel^-}^{\lambda,f\,(AB)}\,
      L_{k_\lambda}(2p_\perp^2/B_f)\, L_{k_\lambda'}(2{p'_\perp}^{\!\!2}/B_f) \ + \nonumber \\
            & & \hspace{-1.1cm}
      8\, F_{kk',q_\parallel^+q_\parallel^-}^{+,f\,(CD)}\,(p_\perp \cdot p'_\perp)\,
      L_{k-1}^1(2p_\perp^2/B_f)\, L_{k'-1}^1(2{p'_\perp}^{\!\!2}/B_f) \bigg]\ ,
  \label{tracess}
  \end{eqnarray}
with
  \begin{equation}
    F_{kk',q_\parallel^+q_\parallel^-}^{\lambda,f\,(XY)} \ = \ \hat X^{\lambda,f}_{k,q_\parallel^+}
      \,\hat X^{\lambda,f}_{k',q_\parallel^-} + (q_\parallel^+ \cdot
      q_\parallel^-)\, \hat Y^{\lambda,f}_{k,q_\parallel^+}\,\hat Y^{\lambda,f}_{k',q_\parallel^-}\ .
  \label{fxy}
  \end{equation}
For simplicity we use the notation $B_f = |q_f B|$.

To work out the integrals in Eq.~(\ref{f0k}) it is convenient to use the
Laguerre-Fourier transforms introduced above. In this way, the integrals
over perpendicular momenta can be performed analytically. As found in
Ref.~\cite{GomezDumm:2017jij}, the expression for $F(0,-m_{\pi^0}^2)$ is
\begin{equation}
F(0,-m_{\pi^0}^2)
 =  -\dfrac{N_C}{\pi}\!\sum_{f=u,d}
                B_f\sum_{k=0}^\infty \int_{q_\parallel}
                \sum_{\lambda = \pm} g_{k,q_\parallel}^{\lambda,f}\Big[
                g_{k,q_\parallel}^{\lambda,f}
                F_{kk,q_\parallel^+ q_\parallel^-}^{\lambda,f\,(AB)} +
                2kB_f\, g_{k,q_\parallel}^{-\lambda,f}
                F_{kk,q_\parallel^+ q_\parallel^-}^{\lambda,f\,(CD)}\Big]\bigg|_{t_\parallel^2 = -m_{\pi^0}^2}
                \ .
\label{f0m2}
\end{equation}
The normalization constant $Z_\parallel^{-1}$ can be obtained by derivation
on the r.h.s.~with respect to $t_\parallel^2$.

To obtain an expression for $Z_\perp^{-1}$ one has to expand
$F(t_\perp^2,t_\parallel^2)$ up to first order in $t_\perp^2$. The
calculation of the corresponding integrals over perpendicular momenta is
sketched in Appendix A. One finally gets
\begin{eqnarray}
\hspace{-0.7cm} Z_{\perp}^{-1} &=& \dfrac{N_C}{4\pi}\, \sum_{f=u,d}\
                \sum_{k=0}^\infty \int_{q_\parallel} \
                \sum_{\lambda = \pm}\, \bigg\{ \Big(
                g_{k,q_\parallel}^{\lambda,f}\,
                F_{kk,q_\parallel^+ q_\parallel^-}^{\lambda,f\,(AB)}\, +
                2kB_f\, g_{k,q_\parallel}^{-\lambda,f}\,
                F_{kk,q_\parallel^+ q_\parallel^-}^{\lambda,f\,(CD)}\Big)
                \times \nonumber \\
            & &  \Big[ k_\lambda \Big(g_{k-1,q_\parallel}^{\lambda,f}+
                   g_{k,q_\parallel}^{\lambda,f}\Big)+ (k_\lambda+1) \Big(g_{k,q_\parallel}^{\lambda,f}+
                   g_{k+1,q_\parallel}^{\lambda,f}\Big)\Big] - \Big(g_{k-1,q_\parallel}^{\lambda,f}+
                   g_{k,q_\parallel}^{\lambda,f}\Big)\, \times
                   \nonumber \\
            & &    \Big[ k_\lambda \Big(g_{k-1,q_\parallel}^{\lambda,f}+
                   g_{k,q_\parallel}^{\lambda,f}\Big)
                   F_{kk-1,q_\parallel^+ q_\parallel^-}^{\lambda,f\,(AB)}\,+
                   \,2B_f k(k-1)\Big(g_{k-1,q_\parallel}^{-\lambda,f}+
                   g_{k,q_\parallel}^{-\lambda,f}\Big)\,
                   F_{kk-1,q_\parallel^+
                   q_\parallel^-}^{\lambda,f\,(CD)}\,\Big]\bigg\}\ .
\label{zperp}
\end{eqnarray}

\subsection{$\pi^0$ decay form factors}

The $\pi^0$-to-vacuum amplitudes for vector and axial vector
quark currents are given by
\begin{eqnarray}
H_\mu^{V,0} (x,\vec p) & = & \langle 0 |\bar\psi(x)\,\gamma_\mu
\frac{\tau^3}{2}\,\psi(x)|\pi_0(\vec p)\rangle \ , \nonumber\\
H_\mu^{A,0} (x,\vec p) & = & \langle 0 |\bar\psi(x)\,\gamma_\mu\gamma_5
\frac{\tau^3}{2}\,\psi(x)|\pi_0(\vec p)\rangle\ .
\end{eqnarray}
As discussed in Ref.~\cite{Coppola:2018ygv}, in the presence of an external
magnetic field these currents can be written in terms of three form factors.
Following the notation in Ref.~\cite{Coppola:2019uyr}, in Euclidean space we
have
\begin{eqnarray}
H_4^{V,0}(x,\vec p)\pm H_3^{V,0}(x,\vec p) & = & \mp\, f^{(V)}_{\pi^0}(p_4\mp p_3)\, e^{ip\,\cdot x} \ , \nonumber\\
H_1^{V,0}(x,\vec p)\pm i H_2^{V,0}(x,\vec p) & = & 0 \ , \nonumber\\
H_4^{A,0}(x,\vec p)\pm H_3^{A,0}(x,\vec p) & = & -i f^{(A\parallel)}_{\pi^0}(p_4\pm p_3)\, e^{ip\,\cdot x} \ , \nonumber\\
H_1^{A,0}(x,\vec p)\pm i H_2^{A,0}(x,\vec p) & = & -i f^{(A\perp)}_{\pi^0}(p_1\pm i p_2)\, e^{ip\,\cdot x} \ .
\label{fdefs}
\end{eqnarray}
If we write the corresponding piece of the bosonic action as
\begin{equation}
S_{\rm bos}\big|_{W^3\,\delta\pi_3} \ = \ \sum_{C=V,A}
\int_{t_\parallel t_\perp}
F^{C}_\mu(t) \, W^{C,3}_\mu(t)\, \delta\pi_3(-t) \ ,
\end{equation}
it is easily seen that
\begin{eqnarray}
    f^{(V)}_{\pi^0} &=& \frac{Z_\parallel^{1/2}}{t^2_\parallel} \,
            \Big[t_3 F_4^V(t) - t_4 F_3^V(t)\Big]  \ , \\
    f^{(A\parallel)}_{\pi^0} &=& i\,\frac{Z_\parallel^{1/2}}{t^2_\parallel} \
            t_\parallel  \cdot F_\parallel^A(t) \ , \\
    f^{(A\perp)}_{\pi^0} &=& i\,\frac{Z_\parallel^{1/2}}{t^2_\perp} \
            t_\perp  \cdot F_\perp^A(t) \ .  \label{faperp}
\end{eqnarray}

The functions $F^{C}_\mu(t)$ can be separated into three pieces
$F^{C,(i)}_\mu(t)$ with $i={\rm I, II, III}$, coming from the various
contributions to the effective action, namely
\begin{eqnarray}
    S^{\,\rm I}_{\rm bos}\big|_{W\,\delta\pi}   &=& - {\rm Tr}[ \mathcal{D}_0^{-1} \delta\mathcal{D}_{W,\pi} ] \ ,\\
    S^{\,\rm II}_{\rm bos}\big|_{W\,\delta\pi}  &=& {\rm Tr}[ \mathcal{D}_0^{-1} \delta\mathcal{D}^{(a)}_{W} \mathcal{D}_0^{-1} \delta\mathcal{D}_{\pi} ] \ , \\
    S^{\,\rm III}_{\rm bos}\big|_{W\,\delta\pi} &=& {\rm Tr}[ \mathcal{D}_0^{-1} \delta\mathcal{D}^{(b)}_{W} \mathcal{D}_0^{-1}
    \delta\mathcal{D}_{\pi}] \ .
\end{eqnarray}
The explicit calculation of $F^{C,(i)}_\mu(t)$ leads to
\begin{eqnarray}
    F^{C,{\rm (I)}}_\mu(t) &=& \frac{N_C}{2} \sum_{f=u,d} \int_{q_\parallel
r_\parallel q_\perp r_\perp} \left[ g\left( (q-r/2)^2 \right) - g\left(
(q-r/2+t/2)^2 \right) \right]
            \ \times \nonumber\\
             & & \qquad\qquad\qquad \trmin_D [\tilde S_f(q_\perp,q_\parallel)\,\gamma_5 \, \Gamma^C] \ h_\mu(r,t-r)
\label{funo} \ , \\
      F^{C,{\rm (II)}}_\mu(t) &=& - i \, 8\pi^2 \, N_C  \sum_{f=u,d} \frac{1}{B_f^2} \int_{q_\parallel q_\perp p_\perp p'_\perp} g(q^2) \
      \exp[i2 \varphi(q_\perp,p_\perp,p'_\perp,t_\perp) / (q_f B)] \, \times \nonumber\\
      && \trmin_D \Big[\tilde S_f(p_\perp,q_\parallel^+) \,  \bar \Gamma^C \gamma_\mu \,
      \tilde S_f(p'_\perp,q_\parallel^-) \, \gamma_5\Big]
\label{fdos} \ , \\
    F^{C,{\rm (III)}}_\mu(t) &=& - 8\pi^2 \,\sigma N_C   \sum_{f=u,d} \frac{1}{B_f^2}
    \int_{q_\parallel r_\parallel q_\perp r_\perp p_\perp p'_\perp} g(q^2) \,
    \exp[i2 \varphi(q_\perp,p_\perp,p'_\perp,t_\perp) /(q_f B)] \times \nonumber \\
    && \trmin_D \Big[\tilde S_f(p_\perp,q_\parallel^+) \, \Gamma^C \, \tilde S_f(p'_\perp,q_\parallel^-) \gamma_5\Big]
     \Big\{ g\left( (p_\perp+p'_\perp - q_\perp - r_\perp/2)^2 + (q_\parallel-r_\parallel/2)^2 \right) \! - \nonumber \\
    && g\left( (p_\perp+p'_\perp - q_\perp - r_\perp/2 + t_\perp/2)^2 + (q_\parallel-r_\parallel/2 + t_\parallel/2)^2 \right)
    \Big\}\, h_\mu(r,t-r) \ ,
\label{ftres}
\end{eqnarray}
where
\begin{eqnarray}
    h_\mu(q,t-q) = \int d^4z\ \exp{\left[-i (t-q)z\right]} \int_0^z d\ell_\mu \exp{\left[i t
    \ell\right]}\
\end{eqnarray}
and
\begin{eqnarray}
    \varphi(q_\perp,p_\perp,p'_\perp,t_\perp) = p_2 (q_1-t_1/2) - p'_2 (q_1+t_1/2) -q_1 t_2 - p_2 p'_1 - (1 \leftrightarrow 2 ) \ .
  \end{eqnarray}
As in the case of the calculation of the $\pi^0$ mass and wave function
renormalization, the integrals over transverse momenta can be performed
analytically after Laguerre-Fourier transforming the nonlocal form factor
functions. The steps to be followed in each case are outlined in Appendix B.
In what follows we just quote the results of this rather lengthy
calculation. The form factors are evaluated at the pion pole,
i.e.~$t_\parallel^2 = - m_{\pi^0}^2$, $t_\perp^2 = 0$.

The calculation of $f^{(A\parallel)}_{\pi^0}$ has been previously performed
in Ref.~\cite{GomezDumm:2017jij}, where the contributions from $F^{C,{\rm
(I)}}_\mu(t)$, $F^{C,{\rm (II)}}_\mu(t)$ and $F^{C,{\rm (III)}}_\mu(t)$ are
quoted. Summing all three contributions one has~\cite{GomezDumm:2017jij}
\begin{eqnarray}
   t_\parallel  \cdot F_\parallel^A(t)\Big|_{t_\perp^2=0}  &=& i\, \frac{N_C}{\pi} \sum_{f=u,d} B_f
   \sum_{k=0}^\infty \
      \int_{q_\parallel} \sum_{\lambda=\pm} g^{\lambda,f}_{k,q_{\parallel}}
      \Big( M^{\,\lambda,f}_{k,q_{\parallel}}\, F_{kk,q_\parallel^+ q_\parallel^-}^{\lambda,f\,(AB)}\, +
      \nonumber \\
     & &  \, 2 k B_f  M^{\,-\lambda,f}_{k,q_{\parallel}}\, F_{kk,q_\parallel^+ q_\parallel^-}^{\lambda,f\,(CD)}
      - \hat A^{\lambda,f}_{k,q_\parallel} \Big) \ .
\end{eqnarray}
Taking account this result, and making use of Eq.~(\ref{pimass}) and the gap
equation~(\ref{gapeq}), one arrives at~\cite{GomezDumm:2017jij}
\begin{equation}
   f^{(A\parallel)}_{\pi^0} \ = \ -\, m_c\, Z_\parallel^{1/2}\,\frac{N_C}{\pi\,t_\parallel^2} \sum_{f=u,d} B_f
   \sum_{k=0}^\infty  \int_{q_\parallel} \sum_{\lambda=\pm} g^{\lambda,f}_{k,q_{\parallel}}
      \Big( F_{kk,q_\parallel^+ q_\parallel^-}^{\lambda,f\,(AB)} +
    2 k B_f  \, F_{kk,q_\parallel^+ q_\parallel^-}^{\lambda,f\,(CD)}\Big)\bigg|_{t_\parallel^2 = -m_{\pi^0}^2} \ .
    \label{fAparallel}
\end{equation}

In the case of $f^{(V)}_{\pi^0}$, it is seen that $F^{V,{\rm (I)}}_\mu(t)$
vanish identically, and the contribution from $F^{V,{\rm (III)}}_\mu(t)$ is
zero. From $F^{V,{\rm (II)}}_\mu(t)$ one obtains
\begin{eqnarray}
   f^{(V\parallel)}_{\pi^0}  &=& Z_\parallel^{1/2}\,\frac{N_C}{\pi} \sum_{f=u,d} B_f
   \sum_{k=0}^\infty \
      \int_{q_\parallel} \frac{\big( q^+_\parallel\cdot t_\parallel \big)}{t_\parallel^2}
      \ \times \nonumber \\
    &  &  \sum_{\lambda=\pm} \lambda\; g^{\lambda,f}_{k,q_{\parallel}} \Big(
      \hat A^{\lambda,f}_{k,q^-_\parallel} \hat B^{\lambda,f}_{k,q^+_\parallel}
      \, -\, 2 k B_f \hat C^{\lambda,f}_{k,q^-_\parallel} \hat
      D^{\lambda,f}_{k,q^+_\parallel} \Big) \bigg|_{t_\parallel^2 =
      -m_{\pi^0}^2}\ .
      \label{fV}
\end{eqnarray}

Finally, for $f^{(A\parallel)}_{\pi^0}$ the calculations sketched in
Appendix B lead to
\begin{eqnarray}
      \frac{t_\perp \cdot F^{A,{\rm (I)}}_\perp(t)}{t_\perp^2}\bigg|_{t_\perp^2=0}  &=& i \frac{N_C}{4\pi}
      \sum_{f=u,d} B_f \sum_{k=0}^\infty \sum_{\lambda=\pm} \int_0^1 d\beta\, \beta \int\limits_{q_\parallel}
      \hat A_{k,q_\parallel}^{\lambda,f}\,
      \Big\{ 2 {g'}^{\lambda,f}_{k,q_{\beta\parallel}^+} + \nonumber \\
   & & B_f\,\Big[(k_\lambda + 1)\,\Big( {g''}^{\lambda,f}_{k+1,q_{\beta\parallel}^+} +
      {g''}^{\lambda,f}_{k,q_{\beta\parallel}^+} \Big) +
      k_\lambda\, \Big( {g''}^{\lambda,f}_{k,q_{\beta\parallel}^+} + {g''}^{\lambda,f}_{k-1,q_{\beta\parallel}^+} \Big)
      \Big]\Big\} \ , \\
    \frac{t_\perp \cdot F^{A,{\rm (II)}}_\perp(t)}{t_\perp^2}\bigg|_{t_\perp^2=0}
    &=& i \frac{N_C}{2\pi} \sum_{f=u,d} B_f s_f \sum_{k=0}^\infty \int\limits_{q_\parallel}\sum_{\lambda=\pm}
    \lambda\, \Big[ 2\,k \, g_{k,q_\parallel}^{\lambda,f} \, H_{kk,q_\parallel^+ q_\parallel^-}^{\lambda,f} \ - \ \nonumber\\
    & &   k_\lambda \, \big(g_{k-1,q_\parallel}^{\lambda,f} \, + \,
    g_{k,q_\parallel}^{\lambda,f}\big)\, H_{k_{-\lambda}k_\lambda,q_\parallel^+ q_\parallel^-}^{\lambda,f} \Big] \ , \\
     \frac{t_\perp \cdot F^{A,{\rm (III)}}_\perp(t)}{t_\perp^2}\bigg|_{t_\perp^2=0}
    &=& -i\frac{\bar\sigma\, N_C}{4\pi}
      \sum_{f=u,d} B_f \sum_{k=0}^\infty \int\limits_{q_\parallel} \sum_{\lambda=\pm}  \bigg\{
      \Big( g^{\lambda,f}_{k,q_\parallel} F_{kk,q_\parallel^+q_\parallel^-}^{\lambda,f(AB)}
       + 2 k B_f \,g^{-\lambda,f}_{k,q_\parallel} F_{kk,q_\parallel^+ q_\parallel^-}^{\lambda,f(CD)} \Big) \times
     \nonumber\\
     && \hspace{-3cm} \int\limits_0^1 d\beta\, \beta \bigg[ 2 {g'}^{\lambda,f}_{k,q_{\beta\parallel}^+}  +
      B_f \Big[(k_\lambda + 1)\,\Big( {g''}^{\lambda,f}_{k+1,q_{\beta\parallel}^+} + {g''}^{\lambda,f}_{k,q_{\beta\parallel}^+} \Big) +
      k_\lambda \Big( {g''}^{\lambda,f}_{k,q_{\beta\parallel}^+} + {g''}^{\lambda,f}_{k-1,q_{\beta\parallel}^+} \Big)\Big] \bigg] -
     \nonumber\\
    && \hspace{-3cm} \dfrac{k_\lambda}{2}\Big[
      \big( {g}^{\lambda,f}_{k-1,q_\parallel} + {g}^{\lambda,f}_{k,q_\parallel} \big)
      \Big( F_{k-1k,q_\parallel^+q_\parallel^-}^{\lambda,f(AB)} -
       F_{kk-1,q_\parallel^+q_\parallel^-}^{\lambda,f(AB)} \Big) + 2 k_{-\lambda}\, B_f\, \times \nonumber \\
      && \hspace{-3cm}  \big( {g}^{-\lambda,f}_{k-1,q_\parallel} + {g}^{-\lambda,f}_{k,q_\parallel} \big)
      \Big( F_{k-1k,q_\parallel^+ q_\parallel^-}^{+,f(CD)} - F_{kk-1,q_\parallel^+ q_\parallel^-}^{+,f(CD)} \Big) \Big]
      \int\limits_0^1 \! d\beta \Big( {g'}^{\lambda,f}_{k,q_{\beta\parallel}^+} + {g'}^{\lambda,f}_{k-1,q_{\beta\parallel}^+} \Big) \bigg\} \ ,
\label{fAIII}
\end{eqnarray}
where ${g'}^{\lambda,f}_{k,q_\parallel}$,
${g''}^{\lambda,f}_{k,q_\parallel}$ indicate derivations with respect to
$q_\parallel^2$, and we have defined
\begin{eqnarray}
    q_{\beta\parallel}^+ & = & q_\parallel +\beta\,t_\parallel/2 \ , \nonumber \\
    H^{\lambda,f}_{kk',q_\parallel^+ q_\parallel^-} & = & \hat A^{\lambda,f}_{k,q_\parallel^+}\, \hat C^{\lambda,f}_{k',q_\parallel^-}
          - (q_\parallel^+\cdot q_\parallel^-)\,\hat B^{\lambda,f}_{k,q_\parallel^+}\,
          \hat D^{\lambda,f}_{k',q_\parallel^-} \ .
\end{eqnarray}
Summing these three contributions, and using the relation
\begin{equation}
    \big( {g'}_{k+1,q_{\parallel}}^{\lambda,f} + {g'}_{k,q_{\parallel}}^{\lambda,f} \big) B_f
       \ = \  g_{k+1,q_{\parallel}}^{\lambda,f} - g_{k,q_{\parallel}}^{\lambda,f}
\end{equation}
(which arises from the properties of Laguerre polynomials), we arrive to a
final expression for $f^{(A\perp)}_{\pi^0}$, given by
\begin{eqnarray}
        f^{(A\perp)}_{\pi^0}  &=& Z_\parallel^{1/2}\, \frac{N_C}{4\pi}
      \sum_{f=u,d} \sum_{k=0}^\infty \int_{q_\parallel} \sum_{\lambda=\pm} \bigg\{
      \Big[-\hat A_{k,q_\parallel}^{\lambda,f} + \bar\sigma \Big( g^{\lambda,f}_{k,q_\parallel}
      F_{kk,q_\parallel^+q_\parallel^-}^{\lambda,f(AB)}
      +  2 k B_f g^{-\lambda,f}_{k,q_\parallel} F_{kk,q_\parallel^+ q_\parallel^-}^{\lambda,f(CD)}
        \Big) \Big]\times \nonumber\\
    && \int\limits_0^1 d\beta\, \beta
      \Big[
      (k_\lambda + 1)\,\Big( g^{\lambda,f}_{k+1,q_{\beta\parallel}^+} - g^{\lambda,f}_{k,q_{\beta\parallel}^+} \Big) -
      k_\lambda \,\Big( g^{\lambda,f}_{k,q_{\beta\parallel}^+} - g^{\lambda,f}_{k-1,q_{\beta\parallel}^+} \Big) \Big] -
      \nonumber \\
    &&  2 B_f s_f \, \lambda \,
        \Big[ 2\,k \, g_{k,q_\parallel}^{\lambda,f} \, H_{kk,q_\parallel^+ q_\parallel^-}^{\lambda,f} \ - \
         k_\lambda \, \big(g_{k-1,q_\parallel}^{\lambda,f} \, + \,  g_{k,q_\parallel}^{\lambda,f}\big) \,
         H_{k_{-\lambda}k_\lambda,q_\parallel^+ q_\parallel^-}^{\lambda,f}
       \Big] \ - \nonumber\\
    && \bar\sigma\dfrac{k_\lambda}{2}\Big[
       \big( {g}^{\lambda,f}_{k-1,q_\parallel} + {g}^{\lambda,f}_{k,q_\parallel} \big)
      \Big( F_{k-1k,q_\parallel^+q_\parallel^-}^{\lambda,f(AB)} -
      F_{kk-1,q_\parallel^+q_\parallel^-}^{\lambda,f(AB)}\Big) + 2 k_{-\lambda} B_f \,\times\nonumber \\
    && \big( {g}^{-\lambda,f}_{k-1,q_\parallel} + {g}^{-\lambda,f}_{k,q_\parallel} \big)
      \Big( F_{k-1k,q_\parallel^+ q_\parallel^-}^{\lambda,f(CD)} - F_{kk-1,q_\parallel^+ q_\parallel^-}^{\lambda,f(CD)} \Big) \Big]
      \int\limits^1_0\! d\beta \Big( g^{\lambda,f}_{k,q_{\beta\parallel}^+} - g^{\lambda,f}_{k-1,q_{\beta\parallel}^+} \Big)
      \bigg\}\bigg|_{t_\parallel^2 = -m_{\pi^0}^2} \! .
      \label{fAperp}
\end{eqnarray}

\subsection{Chiral relations}

It is interesting to study the relations involving form factors and
renormalization constants in the chiral limit, $m_c\to 0$. Firstly, taking
into account the expression in Eq.~(\ref{f0m2}), the gap equation
(\ref{gapeq}) and the relation
\begin{equation}
g_{k,q_\parallel}^{\lambda,f}\,F_{kk,q_\parallel q_\parallel}^{\lambda,f\,(AB)} +
2kB_f\, g_{k,q_\parallel}^{-\lambda,f}\, F_{kk,q_\parallel q_\parallel}^{\lambda,f\,(CD)}
\ = \ \frac{1}{\bar\sigma}\Big(\hat A_{k,q_\parallel}^{\lambda,f} - m_c \,\hat
B_{k,q_\parallel}^{\lambda,f}\Big)\ ,
\label{rel1}
\end{equation}
it is seen that
\begin{equation}
F(0,0) \ = \ -\,\frac{1}{G}\, + \, \dfrac{m_c}{\bar\sigma}\,\dfrac{N_C}{\pi}\sum_{f=u,d}
                B_f\sum_{k=0}^\infty \int_{q_\parallel}
                \sum_{\lambda = \pm} g_{k,q_\parallel}^{\lambda,f}\hat
                B_{k,q_\parallel}^{\lambda,f}\ .
\label{rel2}
\end{equation}
Thus, in the limit $m_c\to 0$, the second term on the r.h.s.~vanishes and
from Eq.~(\ref{pimass}) one obtains $m_{\pi^0}=0$, as expected.

The validity of the Goldberger-Treiman relation
\begin{equation}
f_{\pi^0,0}^{(A\parallel)} \ = \ Z_{\parallel,0}^{-1/2}\,\bar\sigma_0
\label{fpicero}
\end{equation}
and the Gell-Mann-Oakes-Renner relation
\begin{equation}
m_c\,\langle \bar uu + \bar dd\rangle_0 \ = \ -\,m_{\pi^0}^2 \, {f_{\pi^0,0}^{(A\parallel)}}^2
\label{gor}
\end{equation}
in the presence of the external magnetic field have been shown in
Ref.~\cite{GomezDumm:2017jij}. In these equations, subindices 0 indicate
that the quantities have to be evaluated in the chiral limit. Now let us
take into account the expression for $f_{\pi^0}^{(A\perp)}$ in
Eq.~(\ref{fAperp}). For $m_c = 0$, $m_{\pi^0} = 0$, it is seen that the
first term into curly brackets is zero owing to Eq.~(\ref{rel1}), while the
last two terms also vanish since $F_{kk',q_\parallel
q_\parallel}^{\lambda,f\,(XY)}$ is symmetric under the exchange between $k$
and $k'$. Moreover, it is easy to see that
\begin{equation}
    H^{\lambda,f}_{kk,q_\parallel q_\parallel} \ = \ \frac{M^{-\lambda,f}_{k,q_\parallel}}
    {\Delta^f_{k,q_\parallel}}\ ,
\end{equation}
{}from which the piece proportional to $H^{\lambda,f}_{kk,q_\parallel
q_\parallel}$ also vanishes. One gets in this way
\begin{eqnarray}
        f^{(A\perp)}_{\pi^0,0}  &=& Z_{\parallel,0}^{1/2}\, \frac{N_C}{2\pi}
      \sum_{f=u,d} B_f s_f \sum_{k=0}^\infty \int_{q_\parallel}
      \sum_{\lambda=\pm} \lambda\,
         k_\lambda \, \big(g_{k-1,q_\parallel,0}^{\lambda,f} \, + \,  g_{k,q_\parallel,0}^{\lambda,f}\big) \,
         H_{k_{-\lambda}k_\lambda,q_\parallel q_\parallel,0}^{\lambda,f}\  .
      \label{fAperpch}
\end{eqnarray}
On the other hand, from Eqs.~(\ref{zperp}) and (\ref{rel1}) it is seen that
in the chiral limit one has
\begin{eqnarray}
\hspace{-0.8cm} Z_{\perp,0}^{-1} &=& \dfrac{N_C}{4\pi}\, \sum_{f=u,d}\
                \sum_{k=0}^\infty \int_{q_\parallel} \
                \sum_{\lambda = \pm}\, k_\lambda \Big(g_{k_\lambda,q_\parallel,0}^{\lambda,f}+
                   g_{k_{-\lambda},q_\parallel,0}^{\lambda,f}\Big) \bigg[ \frac{1}{\bar\sigma_0}\,
                \Big(  \hat A_{k_\lambda,q_\parallel,0}^{\lambda,f} +
                       \hat A_{k_{-\lambda},q_\parallel,0}^{\lambda,f} \Big)
               - \nonumber \\
            & &  \Big(g_{k_\lambda,q_\parallel,0}^{\lambda,f}+
                   g_{k_{-\lambda},q_\parallel,0}^{\lambda,f}\Big)
                   F_{k_\lambda k_{-\lambda},q_\parallel
                   q_\parallel,0}^{\lambda,f\,(AB)}\,-
                   2k_{-\lambda} B_f \Big(g_{k_\lambda,q_\parallel,0}^{-\lambda,f}+
                   g_{k_{-\lambda},q_\parallel,0}^{-\lambda,f}\Big)
                   F_{k_\lambda k_{-\lambda},q_\parallel
                   q_\parallel,0}^{\lambda,f\,(CD)}\bigg]  \ .
\label{zperpch}
\end{eqnarray}
After some algebra, it can be shown that the factor in square brackets is
equal to $2B_f(k_\lambda - k_{-\lambda}) H_{k_{-\lambda}k_\lambda,q_\parallel
q_\parallel,0}^{\lambda,f}/\bar\sigma_0$. Since $k_\lambda - k_{-\lambda} =
s_f\,\lambda$, by comparing with Eq.~(\ref{fAperpch}) one finally gets
\begin{equation}
f_{\pi^0,0}^{(A\perp)} \ = \
Z_{\parallel,0}^{1/2}\,Z_{\perp,0}^{-1}\,\bar\sigma_0\ .
\label{fpiperp}
\end{equation}
Thus, taking into account Eq.~(\ref{fpicero}), one has
\begin{equation}
\frac{f_{\pi^0,0}^{(A\perp)}}{f_{\pi^0,0}^{(A\parallel)}} \ = \
\frac{Z_{\parallel,0}}{Z_{\perp,0}}\ = \ u_{\pi^0,0}^2 \ .
\label{ratio}
\end{equation}
This result has been also found in the framework of the local NJL model in
Ref.~\cite{Coppola:2019uyr} and (using a different notation) in
Ref.~\cite{Fayazbakhsh:2013cha}, where it is obtained from a modified PCAC
relation.

\section{Finite temperature}

In this section we extend the previous analysis to a system at finite
temperature using the standard Matsubara formalism. To describe the
confinement/deconfinement transitions we include a coupling between the
fermions and the Polyakov loop (PL), assuming that the quarks move on a
uniform background color field. This type of interactions have been
previously considered in nonlocal
models~\cite{Contrera:2007wu,Hell:2008cc,Contrera:2009hk,
Carlomagno:2013ona,GomezDumm:2017iex}, as well as in the local
Polyakov-Nambu-Jona-Lasinio (PNJL)
model~\cite{Fukushima:2003fw,Megias:2004hj,Ratti:2005jh,Roessner:2006xn} and
in Polyakov-quark-meson models~\cite{Schaefer:2009ui,Schaefer:2007pw}. The
background field is given by $\phi = ig\delta_{\mu 0} G_a^\mu \lambda^a /
2$, where $G_a^\mu$ are the SU(3) color gauge fields. Working in the
so-called Polyakov gauge the matrix $\phi$ is given a diagonal
representation $\phi = \phi_3 \lambda_3 + \phi_8 \lambda_8$, and the traced
Polyakov loop $\Phi = \frac{1}{3} \tra \exp (i\phi/T)$ can be taken as an
order parameter of the confinement/deconfinement transitions. Since at mean
field $\Phi$ is expected to be real owing to charge conjugation symmetry,
one has $\phi_8 = 0$ and $\Phi =
[1+2\cos(\phi_3/T)]/3$~\cite{Roessner:2006xn}. In addition, we include a
Polyakov-loop potential $\mathcal{U}(\Phi,T)$ that accounts for effective
gauge field self-interactions. The mean field grand canonical thermodynamic
potential of the system per unit volume under the external magnetic field is
given by~\cite{GomezDumm:2017iex}
\begin{eqnarray}
    \Omega_{B,T}^\mfa
    &=& \dfrac{\bar\sigma^2}{2G} - T \sum_{f=u,d} \dfrac{\vert q_f B\vert}{2\pi}\sum^\infty_{n=-\infty} \sum_{c}
        \int\dfrac{dp_3}{2\pi} \ \bigg[ \ln \left(p_{\parallel nc}^{\,2} +
        {M_{0,p_{\parallel nc}}^{s_f , f}}^2\right) +
        \nonumber \\
    & &
        \sum^\infty_{k=1} \ln \left(\Delta_{k,p_{\parallel nc}}^f \right) \bigg] + \mathcal{U}(\Phi,T) \ ,
\end{eqnarray}
where $\Delta_{k,p_{\parallel nc}}^f$ is the function in
Eq.~(\ref{delta}), and we have defined $\vec p_{\parallel nc} = (p_3,
p_{4\,nc})$, with $p_{4\,nc}=(2n+1)\pi T + \phi_c$. The sum over color
indices runs over $c=r,g,b$, and color background fields are
$(\phi_r,\phi_g,\phi_b) = (\phi_3,-\phi_3,0)$.

Since $\Omega_{B,T}^\mfa$ is divergent, it has to be properly regularized.
We take the prescription followed in Ref.~\cite{GomezDumm:2017iex}, in which
one subtracts to $\Omega_{B,T}^\mfa$ the thermodynamic potential of a free
fermion gas, and then adds it in a regularized form. The regularized
potential is given by
\begin{equation}
    \Omega^{\mfa,{\rm reg}}_{B,T} = \Omega^\mfa_{B,T} - \Omega^{\rm free}_{B,T} + \Omega^{\rm
    free,reg}_{B,T}\ .
\end{equation}
In fact, the ``free'' piece keeps the interaction with the magnetic field
and the PL. The explicit expression of $\Omega^{\rm free,reg}_{B,T}$, for
which the Matsubara sums can be performed analytically, can be found in
Ref.~\cite{GomezDumm:2017iex}.

The form of the PL potential is an additional input of the model. In this
work we take a widely used polynomial form based on a Ginzburg-Landau
ansatz, namely~\cite{Ratti:2005jh,Scavenius:2002ru}
\begin{equation}
  \dfrac{\mathcal{U}(\Phi,T)}{T^4} = - \dfrac{b_2(T)}{2} \Phi^2 -
\dfrac{b_3}{3} \Phi^3 +\dfrac{b_4}{4} \Phi^4 \ ,
\end{equation}
where
  \begin{equation}
    b_2(T) = a_0 + a_1\left(\dfrac{T_0}{T}\right) + a_2 \left(\dfrac{T_0}{T}\right)^2 +
         a_3 \left(\dfrac{T_0}{T}\right)^3 .
  \end{equation}
The numerical values for the parameters $a_i$ and $b_i$, which can be
obtained by a fit to pure gauge lattice QCD results, can be found in
Ref.~\cite{Ratti:2005jh}. This potential leads to a first-order phase
transition at $T_0$, which becomes a further parameter of the model. In the
absence of dynamical quarks, form lattice QCD one would expect a deconfimenent
temperature of about $T_0 = 270$~MeV. However, it has been argued that in
the presence of light dynamical quarks this parameter should be reduced. For
definiteness we will take $T_0 \simeq 200$~MeV, as suggested in
Refs.~\cite{Schaefer:2007pw,Schaefer:2009ui}.

The values of $\bar\sigma$ and $\phi_3$ at the mean field level can be found
by minimizing the regularized thermodynamic potential, i.e.~by solving
the coupled equations
\begin{equation}
    \dfrac{\partial \Omega^{\mfa,{\rm reg}}_{B,T}}{\partial \bar\sigma} = 0\ , \qquad\qquad
    \dfrac{\partial \Omega^{\mfa,{\rm reg}}_{B,T}}{\partial \phi_3} = 0\ .
\end{equation}
Finite temperature meson masses and decay constants can be then calculated
from Eqs.~(\ref{pimass}), (\ref{f0m2}), (\ref{fAparallel}), (\ref{fV}) and
(\ref{fAperp}), following the prescription
\begin{equation}
N_C\int_{q_\parallel} F(q_\parallel,t_\parallel) \ \to
\ T\sum^\infty_{n=-\infty} \sum_{c} \int \dfrac{dq_3}{2\pi}\
F(q_{\parallel nc},t_\parallel)
\end{equation}
and taking the external momentum $\vec t_\parallel = (i\,m_{\pi^0}(T),0)$.
Notice that these mass values correspond to spatial ``screening masses'' for
the zeroth bosonic Matsubara mode ($t_4 =0$). The reciprocals
$m_{\pi^0}(T)^{-1}$ can be understood as the persistence lengths of this
mode, in equilibrium with the heat bath.

\section{Numerical results}

To obtain numerical predictions for the behavior of the quantities defined
in the previous sections, it is necessary to specify the model parameters and
the shape of the nonlocal form factor $g(p^2)$. We consider here the
often-used Gaussian function
\begin{equation}
g(p^2) \ = \ \exp(-p^2/\Lambda^2)\ .
\end{equation}
In general, the form factor introduces an energy scale $\Lambda$ that
represents an effective momentum cut-off. This constant has to be taken as a
free parameter of the model, together with the current quark mass $m_c$ and
the coupling constant $G$ in the effective Lagrangian. In the particular
case of the Gaussian form factor one has the advantage that the integral in
Eq.~(\ref{gpk}) can be performed analytically, which leads to a dramatic
reduction of the computer time needed for numerical calculations of meson
masses and form factors.

As in Refs.~\cite{Pagura:2016pwr,GomezDumm:2017iex}, we fix the free
parameters by requiring the model to reproduce the empirical values (for
vanishing external field) of the pion mass and decay constant, as well as
some phenomenologically adequate values of the quark condensate $\langle
\bar ff\rangle$ ($f=u,d$). Some parameter sets, corresponding to different
values of the condensate, can be found in Ref.~\cite{GomezDumm:2017iex}.
Here we take $m_c = 6.5$~MeV, $\Lambda = 678$~MeV and $G\Lambda^2 = 23.66$,
which lead to $\langle \bar ff\rangle = (- 230~\rm{MeV})^3$. This will be
called Set I. As shown in Ref.~\cite{GomezDumm:2017iex}, for this
parametrization the behavior of quark condensates with the magnetic field,
at zero temperature, are found to be in very good agreement with lattice QCD
results. These parameters have also been used in the previous calculation of
$m_{\pi^0}$ and $f_{\pi^0}^{(A\parallel)}$ for nonzero $B$ carried out in
Ref.~\cite{GomezDumm:2017jij}. In order to test the sensitivity of our
results to the parameters we also consider two alternative sets, which
correspond to quark condensates $\langle \bar ff\rangle = (-
220~\rm{MeV})^3$ and $\langle \bar ff\rangle = (- 240~\rm{MeV})^3$ in
vacuum. The latter are denoted as Sets II and III, respectively.
\begin{center}
\begin{figure}[hbt]
\hspace*{-1.3cm}
\includegraphics[width=1.4\textwidth]{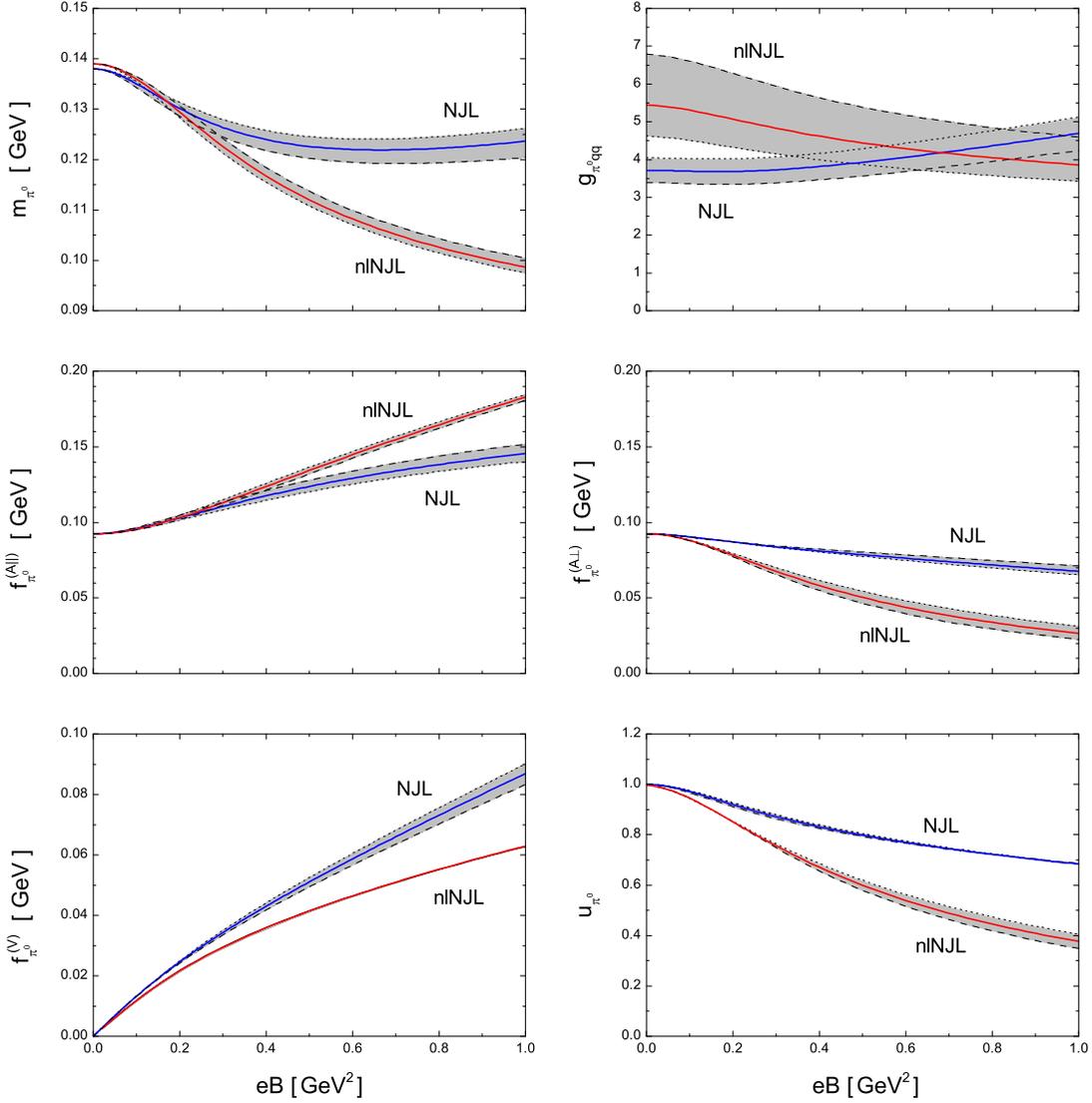}
\caption{Neutral pion properties as functions of $eB$. Solid red lines correspond to Set I, while
the limits of the grey bands correpond to set II (dashed lines) and set III (dotted lines).
Solid blue lines and associated grey bands correspond to local NJL results (see text).}
\label{fig1}
\end{figure}
\end{center}

In Fig.~\ref{fig1} we show our numerical results for various quantities
associated with the neutral pion at zero temperature, as functions of $eB$.
Solid red lines denote the results from Set I, while the limits of the
corresponding grey bands are determined by the results from Set II (dashed
lines) and Set III (dotted lines). For comparison we also include in the
figure the numerical results obtained within the local NJL model, quoted in
Ref.~\cite{Coppola:2019uyr}. Solid blue lines correspond to a
parametrization leading to a constituent quark mass $M=350$~MeV (for $B=0$),
while the limits of the grey bands correspond to $M=320$~MeV (dashed lines)
and $M=380$~MeV (dotted lines). The values of the quark-antiquark
condensates for these parametrizations of the NJL model are $\langle \bar
qq\rangle \simeq (- 243~\rm{MeV})^3$, $(-236~\rm{MeV})^3$ and
$(-250~\rm{MeV})^3$, respectively. It should be noted that the results for
the pion mass and the $f_{\pi^0}^{(A\parallel)}$ form factor have been
previously obtained in Ref.~\cite{GomezDumm:2017jij}.

{}From the graphs in Fig.~\ref{fig1} it can be said that in general our
results do not show a large dependence with the model parametrization. As
shown in the upper left panel in Fig.~\ref{fig1}, the dependence of the
$\pi^0$ mass with the external field for the nonlocal effective model is
significantly stronger than in the case of the local NJL approach. In the
upper right panel of the figure we plot the curves for the effective
coupling constant $g_{\pi^0 q q}$, which shows different behaviors for nlNJL
and NJL models. In the left and right central panels of the figure we quote
the curves corresponding to the axial form factors. Notice that for $B=0$
one has spacial rotation symmetry and both $f_{\pi^0}^{(A\parallel)}$ and
$f_{\pi^0}^{(A\perp)}$ reduce to the usual $\pi^0$ decay constant [see
Eqs.~(\ref{fdefs})]. As the magnetic field increases,
$f_{\pi^0}^{(A\parallel)}$ gets enhanced and $f_{\pi^0}^{(A\perp)}$ gets
reduced. This is in qualitative agreement with the results for the local NJL
model, although for the latter the $B$ dependence is noticeably milder. The
lower left panel shows the behavior of the vector form factor
$f_{\pi^0}^{(V)}$ as a function of $eB$. This form factor is zero at
vanishing external field and shows a monotonic growth with $eB$, with little
dependence on the parametrization. In this case the growth is shown to be
somewhat steeper for the local NJL model. Finally, in the lower right panel
we quote the curves for directional refraction index $u_{\pi^0}$, which is
found to get reduced for increasing external field.
\begin{figure}[hbt]
\includegraphics[width=0.59\textwidth]{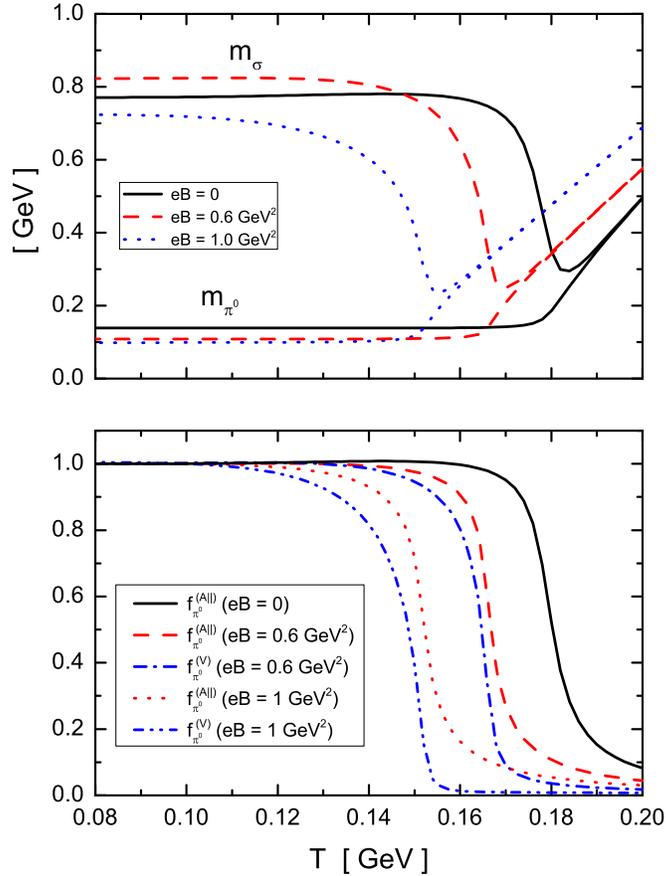}
\caption{$\pi^0$ and $\sigma$ meson masses (upper panel) and normalized
$\pi^0$ decay form factors (lower panel) as functions of the temperature.
Solid lines correspond to $eB = 0$, dashed and dash-dotted lines to
$eB=0.6$~GeV$^2$, and dotted and dash-dot-dotted lines to $eB=1$~GeV$^2$.}
\label{fig2}
\end{figure}

The results for the above discussed quantities, together with those obtained
for $\langle \bar qq\rangle$ (for the analysis of condensates and related
quantities, see Refs.~\cite{Pagura:2016pwr,GomezDumm:2017iex}), allow us to
check the validity of the chiral relations in Eqs.~(\ref{gor}) and
(\ref{ratio}). They are found to be satisfied within 5\% and 0.2\% accuracy,
respectively, for $eB$ up to 1.5~GeV$^2$ (for definiteness, we have
considered parameter Set I). In particular, the opposite behavior of
$f_{\pi^0}^{(A\parallel)}$ and $f_{\pi^0}^{(A\perp)}$ with the magnetic
field can be understood from Eq.~(\ref{ratio}), taking into account that
$u_{\pi^0}$ becomes significantly reduced for increasing $B$. In the NJL, it
has also been shown that the relation $f_{\pi^0}^{(V)} = eB/(8\pi^2
f_{\pi^0}^{(A\parallel)})$ holds in the chiral limit~\cite{Coppola:2019uyr}.
We have checked this relation numerically in the nlNJL model, finding that
it remains only approximately valid (that is, within a 15\% accuracy) for
the chosen range of $eB$.

We turn now to our results for a system al finite temperature. As expected,
at some critical temperature $T_c(B)$ the system undergoes a crossover
transition in which chiral symmetry is partially restored. Moreover, as
shown in Refs.~\cite{Pagura:2016pwr,GomezDumm:2017iex}, this model leads to
inverse magnetic catalysis, in the sense that $T_c$ is found to be a
decreasing function of $B$. This is in agreement with lattice QCD
results~\cite{Bali:2012zg}. It has been also shown that there is a very
small splitting between chiral restoration and deconfinement transition
temperatures, the latter being defined according to the behavior of the
Polyakov loop $\Phi$ (see e.g.~Ref.~\cite{GomezDumm:2017iex} for details).

\begin{figure}[hbt]
\includegraphics[width=0.59\textwidth]{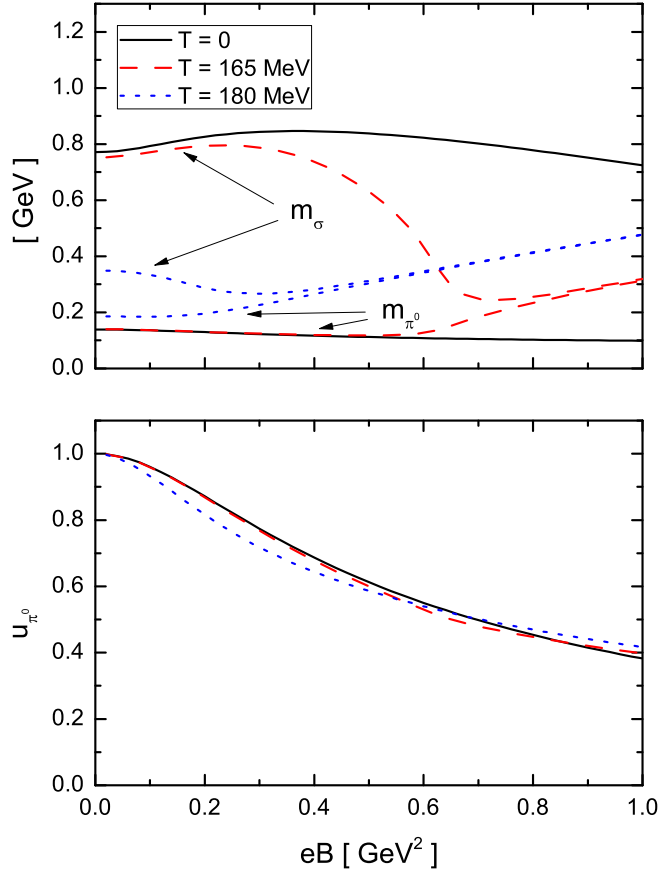}
\caption{$\pi^0$ and $\sigma$ meson masses (upper panel) and directional
refraction index (lower panel) as functions of $eB$, for three
representative values of the temperature.} \label{fig3}
\end{figure}

\begin{figure}[hbt]
\includegraphics[width=0.59\textwidth]{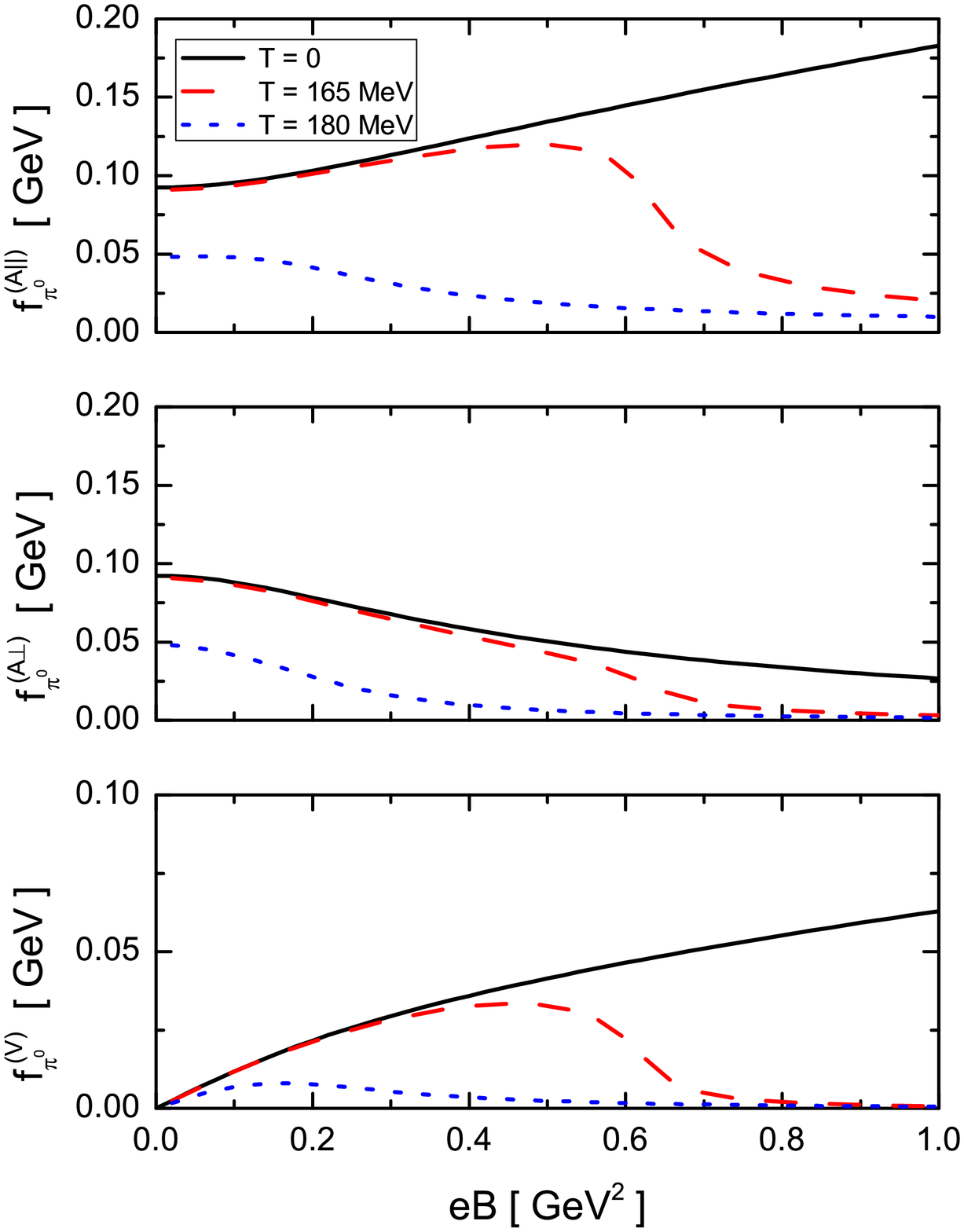}
\caption{From top to bottom, decay form factors $f_{\pi^0}^{(A\parallel)}$,
$f_{\pi^0}^{(A\perp)}$ and $f_{\pi^0}^{(V)}$ as functions of $eB$, for three
representative values of the temperature.} \label{fig4}
\end{figure}

Regarding the quantities we are interested in here, in Fig.~\ref{fig2} we
show the behavior of the $\pi^0$ and $\sigma$ meson masses (upper panel),
and the normalized $\pi^0$ axial and vector decay form factors (lower panel)
as functions of the temperature, for three representative values of the
external magnetic field, namely $eB=0$, $eB=0.6$~GeV$^2$ and $eB=1$~GeV$^2$.
The curves correspond to parameter Set I and a polynomial Polyakov loop
potential, as discussed in Sec.~III. It can be seen that for nonzero $B$ the
masses show a similar qualitative behavior with $T$ as in the $B=0$ case.
The $\pi^0$ mass remains approximately constant up to the critical
temperature, and $\pi^0$ and $\sigma$ masses match above $T_c$, as expected
from chiral symmetry. For large temperatures it is seen that the masses get
steadily increased, the growth being dominated by pure thermal effects. As
stated, the critical temperature gets lower for increasing $B$. The critical
temperatures for the chosen values $eB = 0$, $0.6$~GeV$^2$ and 1~GeV$^2$ are
found to be $T_c \sim 180$~MeV, 165~MeV and 150~MeV,
respectively~\cite{GomezDumm:2017iex}. In the case of the form factors, the
curves for $f_{\pi^0}^{(A\parallel)}$ show a drop at the critical
temperatures, exhibiting once again a qualitatively similar behavior for
zero and nonzero external magnetic field. The curves for
$f_{\pi^0}^{(A\perp)}$, normalized to $f_{\pi^0}^{(A\perp)}(T=0)$, differ
from those of $f_{\pi^0}^{(A\parallel)}$ by less than about 5\%. For clarity
they have not been included in the figure. Finally, the vector form factor
$f_{\pi^0}^{(V)}$ also shows a drop at $T\sim T_c(B)$. The transition in
this case is somewhat steeper than for $f_{\pi^0}^{(A\parallel)}$, and
occurs at a slightly lower temperature. We recall that, at any temperature,
$f_{\pi^0}^{(V)}$ is zero for vanishing external field.

For completeness, in Figs.~\ref{fig3} and~\ref{fig4} we show the behavior of
meson properties as functions of $eB$ for three representative values of the
temperature, namely $T=0$, 165~MeV and 180~MeV. The results for $T=0$, same
as those previously shown in Fig.~\ref{fig1}, are included just for
comparison. The curves for $T=165$~MeV can be understood by looking at the
results in Fig.~\ref{fig2}, which show that this is the critical temperature
that corresponds to $eB\simeq 0.6$~GeV$^2$. Thus, the pion mass and form
factors in Figs.~\ref{fig3} and~\ref{fig4} show approximately the same
behavior as for $T=0$ up to $eB\sim 0.5-0.6$~GeV$^2$. Beyond these values,
as expected from the results in Fig.~\ref{fig2}, one finds an enhancement in
the pion mass and a decrease in the axial and vector form factors. On the
other hand, the curves for $T=180$~MeV are consistent with the fact that the
chiral restoration transition occurs at approximately this temperature for
vanishing magnetic field; the values of the pion mass and axial form factors
are well separated from the $T=0$ values already at $B=0$. Finally, as shown
in the lower panel of Fig.~\ref{fig3}, the behavior of the directional
refraction index $u_{\pi^0}$ is found to be basically independent of the
temperature.

\section{Summary and conclusions}

We have studied the behavior of neutral meson properties in the presence of
a uniform static external magnetic field $B$, in the context of the
so-called nlNJL model. That is, a nonlocal effective approach based on the
Nambu-Jona-Lasinio chiral quark model. Our analysis is a sort of extension
of the work carried out in Ref.~\cite{GomezDumm:2017jij}, where the pion
mass and the decay form factor $f_{\pi^0}^{(A\parallel)}$, at zero
temperature, were studied in this same framework. In the nlNJL model the
effective couplings between quark-antiquark currents include nonlocal form
factors that regularize ultraviolet divergences in quark loop integrals, and
lead to a momentum-dependent effective mass in quark propagators. In order
to obtain closed analytical expressions for meson polarization functions and
pion decay constants in the presence of the external magnetic field, we have
worked out a formalism that involves Ritus transforms of the Dirac fields.

We have firstly concentrated in the analysis at zero temperature of the form
factors associated with pion-to-vacuum matrix elements of the vector and
axial vector hadronic currents. In agreement with the model independent
analysis in Ref.~\cite{Coppola:2018ygv}, it is seen that for nonzero $B$
three nonvanishing independent form factors can be defined. Two of them,
$f_{\pi^0}^{(A\parallel)}$ and $f_{\pi^0}^{(A\perp)}$, correspond to the
axial-vector current, and the other one, $f_{\pi^0}^{(V)}$, to the vector
piece. We have also calculated the neutral pion directional refraction
index, $u_{\pi^0}$, which in general is different from one. In addition,
chiral relations are shown to be valid in the presence of the external
field.

For the numerical calculations we have considered the case of Gaussian
nonlocal form factors, choosing sets of model parameters that were
previously found to reproduce the empirical values of the pion mass and
decay constant at $B=0$, and lead to values of quark condensates at nonzero
$B$ that are compatible with lattice QCD calculations. Taking into account
external fields in the range from zero to $eB = 1$~GeV$^2$, from our numerical
results it is noticed that all studied quantities show a strong dependence with $B$.
As discussed in Sec.~IV, in most cases this dependence is found to be
significantly larger than that observed in the local NJL
model~\cite{Coppola:2019uyr}. On the other hand, it is seen that in general
the results are rather stable under changes in the model parameters. It has
been also verified that chiral relations remain approximately valid for the
chosen parameter sets.

We have extended the calculations to finite temperature systems, including
the couplings of fermions to the Polyakov loop and a PL polynomial potential
that accounts for effective color gauge field self-interactions. As
expected, it is seen that the system undergoes a phase transition
corresponding to the restoration of SU(2) chiral symmetry. The model
predicts the existence of inverse magnetic catalysis, leading to a decrease
of the critical temperature $T_c$ with the magnetic field. Concerning the
behavior of meson masses, it is seen that the $\pi^0$ mass remains
approximately constant up to $T_c(B)$, while the $\sigma$ meson mass begins
to drop earlier. Beyond the transition both masses become degenerate, as
expected, and show a thermal growth for large $T$. Regarding the thermal
behavior of the form factors, we observe that they remain approximately
constant until temperatures close to $T_c(B)$ are reached, and then they
show sudden drops. Finally, the directional refraction index $u_{\pi^0}$ is
found to be basically independent of the temperature. To provide an
alternative view, we have also included some graphs showing the behavior of
the studied quantities as functions of the magnetic field, for some selected
values of the temperature.

\section*{Acknowledgements}

This work has been supported in part by Consejo Nacional de Investigaciones
Cient\'ificas y T\'ecnicas and Agencia Nacional de Promoci\'on Cient\'ifica
y Tecnol\'ogica (Argentina), under Grants No.~PIP17-700 and No.
PICT17-03-0571, respectively, and by the National University of La Plata
(Argentina), Project No.~X824.

\section*{Appendix A}

\newcounter{erasmo}
\renewcommand{\thesection}{\Alph{erasmo}}
\renewcommand{\theequation}{\Alph{erasmo}\arabic{equation}}
\setcounter{erasmo}{1} \setcounter{equation}{0} % \setcounter{table}{0}

We outline here the derivation of the relation in Eq.~(\ref{zperp}), which
can be obtained following a procedure similar to that described in the
appendices of Ref.~\cite{GomezDumm:2017jij}.

It is easy to see that the expression in Eq.~(\ref{f0k}) can be rearranged in
the form
\begin{equation}
F(t_\perp^2,t_\parallel^2) \ = \ -\,128\,\pi^2\,N_C \sum_{f=u,d}\,\frac{1}{B_f^2}
                \sum_{k,k'=0}^\infty \int_{q_\parallel} \ \bigg[
                \sum_{\lambda = \pm}\,F_{kk',q_\parallel^+ q_\parallel^-}^{\lambda,f\,(AB)}\,
                I_{kk',q_\parallel}^{\lambda,f\,(0)} + F_{kk',q_\parallel^+ q_\parallel^-}^{+,f\,(CD)}\,
                I_{kk',q_\parallel}^{f\,(1)}\bigg]\ .
\label{f0kaux}
\end{equation}
Taking the Laguerre-Fourier transforms of the nonlocal form factors given by
Eq.~(\ref{gpk}), and changing the integration variables to dimensionless
vectors $u = -\sqrt{(2/B_f)}\, p_\perp$, $v = \sqrt{(2/B_f)}\, p'_\perp$, $w
= \sqrt{(2/B_f)}\, (p_\perp - q_\perp)$ and $r_\perp = \sqrt{(2/B_f)} \,
t_\perp$ in the plane perpendicular to $\vec B$, the integrals
$I_{kk',q_\parallel}^{\lambda,f\,(0)}$ and $I_{kk',q_\parallel}^{f\,(1)}$
are given by
\begin{eqnarray}
    I_{kk',q_\parallel}^{\lambda,f\,(0)} &=& \frac{B_f^3}{2}\,(-1)^{k+k'}\sum_{m,m'=0}^\infty
        (-1)^{m+m'}\,g_{m,q_\parallel}^{\lambda,f}\,g_{m',q_\parallel}^{\lambda,f}\, K_{kk'mm'}^{\lambda,f\,(0)}\ ,
        \nonumber \\
    I_{kk',q_\parallel}^{f\,(1)} &=& 2\,B_f^4\,(-1)^{k+k'}\sum_{m,m'=0}^\infty
        (-1)^{m+m'-1}\,g_{m,q_\parallel}^{+,f}\,g_{m',q_\parallel}^{-,f}\, K_{kk'mm'}^{f\,(1)}\ ,
\end{eqnarray}
with
\begin{eqnarray}
    K_{kk'mm'}^{\lambda,f\,(0)} \!
      &=& \!\int\limits_{u\, v\, w} \!\!\!\exp\!\big[-w^2\big]\exp\!\big[-u^2\!-u\cdot w + is_f(u_1w_2-u_2w_1)\big]\,
          L_{k_\lambda}(u^2)\,L_{m_\lambda}\big[(u+w)^2\big]\!\times \nonumber \\
      & & \qquad\exp\big[-v^2\!-v\cdot w +
          is_f(v_1w_2-v_2w_1)\big]\,L_{k'_\lambda}(v^2)L_{m'_\lambda}\big[(v+w)^2\big]\times \nonumber\\
      & & \qquad\exp\left\{
      \dfrac{is_f}{2}\big[r_1(2w_2+u_2+v_2)-r_2(2w_1+u_1+v_1)\big]\right\} \ ,
          \nonumber\\
    K_{kk'mm'}^{f\,(1)} \!
      &=& \! - \!\!\int\limits_{u\, v\, w} \!\!\!\exp\!\big[-w^2\big]\exp\!\big[-u^2\!-u\cdot w + is_f(u_1w_2-u_2w_1)\big]\,
          L_{k-1}^1(u^2)\,L_{m_+}\big[(u+w)^2\big]\!\times \nonumber \\
      & & \qquad (u\cdot v) \, \exp\big[-v^2\!-v\cdot w + is_f(v_1w_2-v_2w_1)]\,L_{k'-1}(v^2)L_{m'_-}[(v+w)^2\big]\times \nonumber\\
      & & \qquad\exp\left\{
      \dfrac{is_f}{2}\big[r_1(2w_2+u_2+v_2)-r_2(2w_1+u_1+v_1)\big]\right\} \ .
\end{eqnarray}
Notice that $K_{kk'mm'}^{\lambda,f\,(0)}$ and $K_{kk'mm'}^{f\,(1)}$ do not
depend on the magnetic field, but they do depend on the external momenta
$r_\perp=(r_1,r_2)$. Thanks to rotational symmetry we can chose $r_1=r$ and
$r_2=0$. Using the polar coordinates
\begin{eqnarray}
u_1 &=& u \cos(\alpha-\beta)\ , \qquad \qquad v_1 = v \cos(\alpha-\gamma)\ ,
 \qquad \qquad w_1 = w \cos(\alpha) \ ,  \nonumber \\
u_2 &=& u \sin(\alpha-\beta)\ , \qquad \qquad v_2 = v \sin(\alpha-\gamma)\ ,
\qquad \qquad w_2 = w \sin(\alpha)\ ,
\end{eqnarray}
and performing a series expansion around $r=0$ in the exponential, we can
integrate the variable $\alpha$. As seen from Eq.~(\ref{defzperp}), only the
terms quadratic in the external momenta $t_\perp^2$ will contribute to the
perpendicular renormalization constant. Thus, we have
\begin{eqnarray}
   \int\limits_0^{2\pi} d\alpha \,
          \exp\left\{is_fr \left[2w\sin\alpha+u\sin(\alpha-\beta)+v\sin(\alpha-\gamma)\right]/2 \right\}
          & = & \nonumber \\
      & & \hspace{-12cm} 2\pi - \dfrac{\pi r^2}{8} \left[u^2 + v^2 + 4w^2 + 4uw\cos\beta + 4vw\cos\gamma
          + 2uv\cos(\beta-\gamma)\right] + \mathcal{O}(r^4)\ .
\end{eqnarray}
The calculation of the remaining integrals can be performed with the aid of
the useful relations
\begin{eqnarray}
\!\!\!
\frac{1}{2\pi}\int_0^{2\pi}\!\! d\theta\; L_n(x^2\!+y^2\!+2 xy \cos\theta) \,\exp[-xy\exp(\pm i\theta)]
& = & L_n(x^2)\,L_n(y^2)\ ,
\label{master} \\
\!\!\!\frac{1}{2\pi}\int_0^{2\pi}\!\! d\theta\;\cos\theta\, L_n(x^2\!+y^2\!+2 xy \cos\theta)
\,\exp[-xy\exp(\pm i\theta)]
& = & -\,\frac{xy}{2}\Big[\frac{L_n^1(x^2)\,L_n^1(y^2)}{n+1}+\nonumber \\
& & \frac{L_{n-1}^1(x^2)\,L_{n-1}^1(y^2)}{n}\Big]\ ,
\label{master2} \\
\!\!\!\frac{1}{2\pi}\int_0^{2\pi}\!\! d\theta\;\sin\theta\, L_n(x^2\!+y^2\!+2 xy \cos\theta)
\,\exp[-xy\exp(\pm i\theta)]
& = & \mp\,\frac{ixy}{2}\Big[\frac{L_n^1(x^2)\,L_n^1(y^2)}{n+1}-\nonumber \\
& & \frac{L_{n-1}^1(x^2)\,L_{n-1}^1(y^2)}{n}\Big]\ ,
\label{master3}
\end{eqnarray}
together with the orthogonality properties of the generalized Laguerre
polynomials. This leads to
\begin{eqnarray}
  I_{kk',q_\parallel}^{\lambda,f\,(0)} &=& \frac{B_f^3}{128\,\pi^3}\Bigg\{ \;g_{k,q_\parallel}^{\lambda,f}
                       \,g_{k,q_\parallel}^{\lambda,f}\,\delta_{kk'} \ - \
                       \frac{t_\perp^2}{4B_f}\; \bigg[ \Big[ (2k_\lambda +1)\, g_{k,q_\parallel}^{\lambda,f}
                       \,g_{k,q_\parallel}^{\lambda,f}\, +\, (k_\lambda +1)\, g_{k,q_\parallel}^{\lambda,f}
                       \,g_{k+1,q_\parallel}^{\lambda,f}\, +\nonumber\\
                    & & k_\lambda \, g_{k,q_\parallel}^{\lambda,f}
                       \,g_{k-1,q_\parallel}^{\lambda,f}\Big]\, \delta_{kk'}
                       - \ \dfrac{k_\lambda+1}{2}\;\big(g_{k,q_\parallel}^{\lambda,f} \,
                        + \, g_{k+1,q_\parallel}^{\lambda,f} \big)^2 \,\delta_{kk'-1} \ - \nonumber\\
                    & & \dfrac{k_\lambda}{2}\;  \big(g_{k,q_\parallel}^{\lambda,f} \,
                        + g_{k-1,q_\parallel}^{\lambda,f} \big)^2 \,\delta_{kk'+1} \bigg] \
                        + \mathcal{O}(r^4) \Bigg\}
  \label{i0}
  \end{eqnarray}
and
\begin{eqnarray}
  I_{kk',q_\parallel}^{f\,(1)} &=& \frac{k\, B_f^4}{32\pi^3} \Bigg\{\;g_{k,q_\parallel}^{+,f}
                   \,g_{k,q_\parallel}^{-,f}\,\delta_{kk'} \ + %\nonumber\\
                     %& &
                     \frac{t_\perp^2}{8\, B_f}\; \bigg[ \Big[
                     - \ s_f \, \big(g_{k-1,q_\parallel}^{+,f} \,g_{k+1,q_\parallel}^{+,f}\, -
                               \, g_{k-1,q_\parallel}^{-,f} \,g_{k+1,q_\parallel}^{-,f}\big) \nonumber\\
                & & -\, k \, \big( 4\,
                       g_{k,q_\parallel}^{+,f}\,g_{k,q_\parallel}^{-,f}\,
                      +\,  g_{k-1,q_\parallel}^{+,f} \,g_{k,q_\parallel}^{-,f}\,
                      +\,  g_{k,q_\parallel}^{+,f}\,g_{k+1,q_\parallel}^{-,f}\,
                      +\,  g_{k+1,q_\parallel}^{+,f} \,g_{k,q_\parallel}^{-,f}\,
                      +\,  g_{k,q_\parallel}^{+,f}\,g_{k-1,q_\parallel}^{-,f} \big)
                               \Big]\, \delta_{kk'} + \nonumber\\
                    & &
                  (k+1)\; \big(g_{k,q_\parallel}^{+,f} \, + \, g_{k+1,q_\parallel}^{+,f} \big) \;
                      \big(g_{k,q_\parallel}^{-,f} \, + \, g_{k+1,q_\parallel}^{-,f} \big)
                      \,\delta_{kk'-1} \ + \nonumber\\
                    & &
                  (k-1)\; \big(g_{k-1,q_\parallel}^{+,f} \, + \, g_{k,q_\parallel}^{+,f} \big) \;
                          \big(g_{k-1,q_\parallel}^{-,f} \, + \, g_{k,q_\parallel}^{-,f} \big)
                          \,\delta_{kk'+1} \bigg] \
                        + \   \mathcal{O}(r^4) \Bigg\} \ .
\label{i1}
\end{eqnarray}
Replacing the results in Eq.~(\ref{i0}) and (\ref{i1}) into
Eq.~(\ref{f0kaux}) one arrives at our final expression, quoted in
Eq.~(\ref{zperp}).

\section*{Appendix B}

\newcounter{erasmo2}
\renewcommand{\thesection}{\Alph{erasmo2}}
\renewcommand{\theequation}{\Alph{erasmo2}\arabic{equation}}
\setcounter{erasmo2}{2} \setcounter{equation}{0} % \setcounter{table}{0}

To calculate the contributions $t_\perp \cdot F^{A,{\rm (I)}}_\perp$ and
$t_\perp \cdot F^{A,{\rm (III)}}_\perp$ to $f^{(A\perp)}_{\pi^0}$ we start
by integrating $t_\perp \cdot h_\perp$ along the straight line path
$\ell_\mu = \beta z_\mu$. One has
\begin{equation}
      t_\perp \cdot h_\perp = -i (2\pi)^4 \int_0^1 d\beta \
                \delta^{(2)} \left( r_\parallel - (1-\beta) t_\parallel \right) \
      \partial_\beta  \delta^{(2)} \left( r_\perp - (1-\beta) t_\perp
      \right)\ .
      \label{hh}
\end{equation}
Given the definition in Eq.~(\ref{faperp}), we perform a series expansion
around $t_\perp = 0$ up to order $t_\perp^2$ for each contribution to the
axial perpendicular decay constant, similar to the case of $Z_\perp$. Thus,
we find
\begin{eqnarray}
      t_\perp \cdot F^{A,{\rm (I)}}_\perp(t) &=& \frac{i\,N_C}{2}\, t_\perp^2 \sum_{f=u,d} \int_q \trmin_D \Big[\tilde S_f(q)\Big]
          \int_0^1 d\beta \ \beta \
      \Big[ g'(q_\perp^2+q_{\beta\parallel}^{+2}) \ + \ q_\perp^2\, g''(q_\perp^2+q_{\beta\parallel}^{+2}) \Big] \ , \nonumber\\
      && \\
      t_\perp \cdot F^{A,{\rm (II)}}_\perp(t) &=& - 8\pi^2 \, N_C  \sum_{f=u,d} \frac{s_f}{B_f^3} \int_{p_\perp p'_\perp q} g(q^2) \,
      \exp\big[i2 \varphi_0(q_\perp,p_\perp,p'_\perp) / (q_f B)\big] \, \times \nonumber\\
      && \trmin_D \Big[\tilde S_f(p_\perp,q_\parallel^+) \,  \gamma_5 \, (t_\perp \cdot \gamma_\mu ) \,
      \tilde S_f(p'_\perp,q_\parallel^-) \, \gamma_5\Big] \, |t_\perp \times (q_\perp-u_\perp)|\ ,\\
      t_\perp \cdot F^{A,{\rm (III)}}_\perp(t) &=& -i \, 4\pi^2 \, N_C \,\sigma  \sum_{f=u,d} \frac{1}{B_f^2}
        \int_{p_\perp p'_\perp q}  g(q^2) \,  \exp\big[i2\varphi_0(q_\perp,p_\perp,p'_\perp)/(q_f B)\big]
         \ \times \nonumber \\
    &&
    \trmin_D \Big[\tilde S_f(p_\perp,q_\parallel^+) \, \gamma_5 \,
    \tilde S_f (p'_\perp,q_\parallel^-) \, \gamma_5\Big] \,
    \times \sum_{i=1}^3 f_i(q_\perp, u_\perp, t_\perp)\ ,
    %\big[f_1(q_\perp, u_\perp, t_\perp)  +  f_2(q_\perp, u_\perp, t_\perp)  +
    %\, f_3(q_\perp, u_\perp, t_\perp) \big] \, ,
\end{eqnarray}
where
\begin{eqnarray}
    f_1 &=& - \frac{2\, i }{B_f} \, \big| t_\perp \times (q_\perp - u_\perp) \big| \;
        (u_\perp \cdot t_\perp)
        \int\limits_0^1 d\beta \ \left[ g'(u_\perp^2 + q_{\beta\parallel}^{-2}) \
        - \ g'(u_\perp^2+q_{\beta\parallel}^{+2}) \right] \ , \\
     f_2 &=&  2 \, \big( u_\perp \cdot t_\perp \big)^2
          \int\limits_0^1 d\beta  \left[ (1-\beta)\, g''(u_\perp^2 + q_{\beta\parallel}^{-2}) \
          + \ \beta\, g''(u_\perp^2 + q_{\beta\parallel}^{+2}) \right] \ , \\
     f_3 &=&  t_\perp^2
          \int\limits_0^1 d\beta \ \left[ (1-\beta)\, g'(u_\perp^2 + q_{\beta\parallel}^{-2}) \
          + \ \beta \, g'(u_\perp^2 + q_{\beta\parallel}^{+2}) \right] \ .
\end{eqnarray}
In these expressions we use the notation $q_{\beta\parallel}^+ = q_\parallel +\beta\,t_\parallel/2$,
$q_{\beta\parallel}^- = q_\parallel - (1-\beta)\,t_\parallel/2$, $u_\perp = p_\perp + p'_\perp - q_\perp$,
and $\varphi_0(q_\perp,p_\perp,p'_\perp) =
\varphi(q_\perp,p_\perp,p'_\perp,0)$.

To calculate the integrals over perpendicular momenta we follow a similar
procedure as that described in App.~A. That is, we introduce the
Laguerre-Fourier transforms of the form factors and the expressions for the
traces. Afterwards, performing appropriate changes of the integration
variables, the integrals can be calculated using the orthogonality
properties of the generalized Laguerre polynomials, the properties in
Eqs.~(\ref{master}), (\ref{master2}) and (\ref{master3}), and the relations
\begin{eqnarray}
 \!\!\!
  \frac{1}{2\pi}\int\limits_0^{2\pi}\!\! d\theta\;\cos(2\theta)\, L_n(x^2\!+y^2\!+2 xy \cos\theta)\,\exp[-xy\, e^{\pm i\theta}]
    &=& \frac{1}{2}\Big[F_n(x,y)+F_{n-2}(x,y) \Big] \ , \ \ \ \
    \label{master4} \\
  \!\!\!
  \frac{1}{2\pi}\int\limits_0^{2\pi}\!\! d\theta\;\sin(2\theta)\, L_n(x^2\!+y^2\!+2 xy \cos\theta)\,\exp[-xy\, e^{i\pm \theta}]
    &=& \frac{\pm i}{2}\Big[F_n(x,y)-F_{n-2}(x,y) \Big] \ ,  \ \ \ \
    \label{master5}
\end{eqnarray}
 and
\begin{eqnarray}
  \!\!\!
  \int\limits_0^{\infty}\!\! dw^2 e^{-w^2} \Big[F_m(x,w) \pm F_{m-2}(x,w)\Big] \Big[F_n(y,w) \pm F_{n-2}(y,w)\Big] &=& \nonumber\\
    & & \hspace{-7cm} \big(\delta_{mn} \pm \delta_{mn-2} \big)F_m(x,y) \ + \
        \big(\delta_{mn} \pm \delta_{mn+2} \big)F_{m-2}(x,y) \ , \ \ \ \
    \label{master6} \\
  \!\!\!
  \int\limits_0^{\infty}\!\! dw^2 e^{-w^2} \Big[F_m(x,w) \pm F_{m-2}(x,w)\Big] \Big[F_n(y,w) \mp F_{n-2}(y,w)\Big] &=& \nonumber\\
    & & \hspace{-7cm} \big(\delta_{mn} \mp \delta_{mn-2} \big)F_m(x,y) \ - \
        \big(\delta_{mn} \mp \delta_{mn+2} \big)F_{m-2}(x,y) \ , \ \ \ \
    \label{master7}
\end{eqnarray}
where
\begin{equation}
    F_n(x,y) = L_{n+1}(x^2)L_{n+1}(y^2)-\dfrac{1}{n+2}L_{n+1}^1(x^2)L_{n+1}^1(y^2) +
    \dfrac{1}{n+1}L_n^1(x^2)L_n^1(y^2)\ .
\end{equation}
In the case of calculations of $t_\perp \cdot F^{A,{\rm (II)}}_\perp(t)$,
and $f^{(V)}_{\pi^0}$, some relations between the Bessel functions
$J_\nu(x)$ and the Laguerre polynomials are also required (see appendices in
Ref.~\cite{GomezDumm:2017jij}).

In this way, after a lengthy calculation, one arrives at the expression for
$f^{(A\perp)}_{\pi^0}$ in Eq.~(\ref{fAperp}).

\end{document}